\documentclass[aps,prb,superscriptaddress,twocolumn,floatfix,showpacs,citeautoscript,longbibliography,hyperlinks]{revtex4-1}

\usepackage{amsmath}
\usepackage{amssymb}
\usepackage{amsfonts}
\usepackage{dsfont}
\usepackage{graphicx,graphics}
\usepackage{color}
\usepackage[autostyle]{csquotes}
\usepackage[colorlinks=true,citecolor=blue]{hyperref}

\newcommand{\ket}[1]{\lvert #1\rangle}
\newcommand{\bra}[1]{\langle#1 \rvert}

\definecolor{ThesisBlue}{RGB}{40,140,170}
\definecolor{CPSRed}{RGB}{170,40,40}
\definecolor{Grey}{RGB}{100,100,100}
\definecolor{Grey2}{RGB}{150,150,150}
\newcommand\scalemath[2]{\scalebox{#1}{\mbox{\ensuremath{\displaystyle #2}}}}

\begin{document}

\title{Noise and full counting statistics of a Cooper pair splitter}
\author{Nicklas Walldorf}
\affiliation{Center for Nanostructured Graphene (CNG), Department of Physics, Technical University of Denmark, DK-2800 Kongens Lyngby, Denmark}
\author{Fredrik Brange}
\affiliation{Department of Applied Physics, Aalto University, 00076 Aalto, Finland}
\author{Ciprian Padurariu}
\affiliation{Institute for Complex Quantum Systems and IQST, University of Ulm, 89069 Ulm, Germany}
\author{Christian Flindt}
\affiliation{Department of Applied Physics, Aalto University, 00076 Aalto, Finland}

\date{\today}

\begin{abstract}
We investigate theoretically the noise and the full counting statistics of electrons that are emitted from a superconductor into two spatially separated quantum dots by the splitting of Cooper pairs and further on collected in two normal-state electrodes. With negatively-biased drain electrodes and a large superconducting gap, the dynamics of the Cooper pair splitter can be described by a Markovian quantum master equation.  Using techniques from full counting statistics, we evaluate the electrical currents, their noise power spectra, and the power-power correlations in the output leads. The current fluctuations can be attributed to the competition between Cooper pair splitting and elastic cotunneling between the quantum dots via the superconductor. In one regime, these processes can be clearly distinguished in the cross-correlation spectrum with peaks and dips appearing at  characteristic frequencies associated with elastic cotunneling and Cooper pair splitting, respectively. We corroborate this interpretation by analyzing the charge transport fluctuations in the time domain, specifically by investigating the $g^{(2)}$-function of the output currents. Our work identifies several experimental signatures of the fundamental transport processes involved in Cooper pair splitting and provides specific means to quantify their relative strengths. As such, our results may help guide and interpret future experiments on current fluctuations in Cooper pair splitters.  
\end{abstract}

\maketitle

\section{Introduction}

\begin{figure}[b!]
  \centering
  \includegraphics[width=0.95\columnwidth]{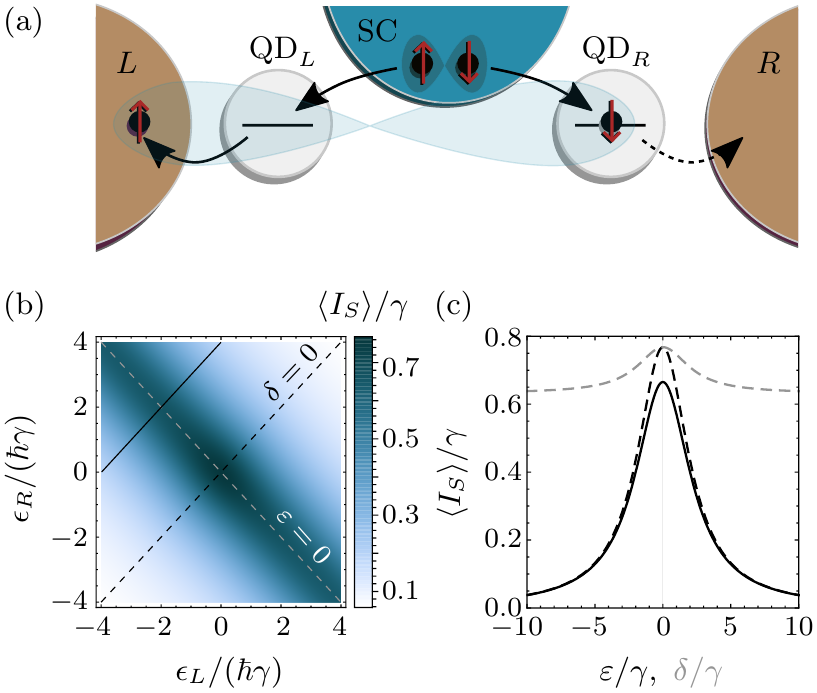}
  \caption{Cooper pair splitter and average current. (a) Cooper pairs from a superconducting lead are split into two spin-entangled electrons that tunnel into separate quantum dots before reaching the output drains. (b) Average current from the superconductor into the drain electrodes as a function of the quantum dot levels. (c) The average current as a function of the sum, $\varepsilon=(\epsilon_L+\epsilon_R)/\hbar$, and the detuning, $\delta=(\epsilon_L-\epsilon_R)/\hbar$, of the quantum dot levels corresponding to the three cuts in the left panel. The results are obtained from Eq.~(\ref{eq:av_curr}) with the parameters $\gamma_{\text{CPS}}=\gamma_\text{EC}=\gamma$, $\gamma_L=1.5\gamma$, $\gamma_R=0.5\gamma$.}
\label{fig:setup}
\end{figure}

Superconductors can serve as sources of entanglement in solid-state quantum circuits.\cite{Lesovik2001,Recher:Andreev} Electrons in the superconductor are paired up in spin-entangled states and by splitting these Cooper pairs, entanglement between distant electrons may be achieved. Specifically, electrons from a Cooper pair may tunnel into different normal-state electrodes, while preserving the entanglement of their spins. The process can be enhanced by using quantum dots with strong Coulomb interactions, which prevent electron pairs from tunneling into the same output lead, see Fig.~\ref{fig:setup}. To certify the entanglement of the split Cooper pairs, it has been suggested that  Bell inequalities can be formulated for the cross-correlations of the output currents, using ferromagnetic leads as spin filters.\cite{Kawabata2001,Cht2002,Sauret:Spin,Malkoc:Full,Soller:Generic,Brange2017}

Following the theoretical proposals to generate non-local entanglement using Cooper pair splitters,\cite{Lesovik2001,Recher:Andreev} several experiments have realized these ideas in practice. Cooper pair splitters have been implemented in a variety of superconductor hybrid systems, \cite{Beckmann2004,Russo2005,Deacon:Cooper,Bruhat2018} some of which employ InAs nanowires \cite{Das2012,Hofstetter:Cooper,Hofstetter2011,Fulop2015,Ueda2019}, carbon nanotubes \cite{Herrmann:Carbon,Herrmann:Spectroscopy,Schindele:Near,Fueloep:Local,Schindele:Nonlocal}, or graphene-based nanostructures.\cite{Tan:Cooper,Borzenets:High} The Cooper pair splitters can be characterized by measuring the conductance\cite{Hofstetter:Cooper,Hofstetter2011,Fulop2015,Ueda2019,Herrmann:Carbon,Herrmann:Spectroscopy,Schindele:Near,Fueloep:Local,Schindele:Nonlocal,Tan:Cooper} or the noise\cite{Das2012}, and the splitting efficiency is in some cases approaching unity,\cite{Schindele:Near,Borzenets:High} with one of the main competing processes being elastic cotunneling between the dots.\cite{Hofstetter2011,Tan:Cooper} With this experimental progress, one may hope that Cooper pair splitters can soon be integrated into larger quantum circuits, aiming for solid-state quantum information processing.  

On the theory side, Cooper pair splitters can be described using a variety of techniques depending on the specific device architecture and the operating conditions. For non-interacting systems, tight-binding models\cite{Burset:Microscopic,Braunecker2013} or scattering theory\cite{Sadovskyy:Unitary} provide a convenient theoretical framework.
Interactions can be included using 
Green function techniques,\cite{Melin2008,Chevallier:Current,rech2012a,Burset2016,celis2017a}
quantum master equations,\cite{Sauret:Spin,Sauret:Quantum,Konig,Amitai2016,Hussein2017,Islam2017,Sanchez2018}
or the real-time diagrammatic approach to quantum transport.\cite{Hiltscher2011,Sothmann2014,Trocha2015,wrze2017a,trocha2018a} In most cases, these methods enable numerical calculations of the average currents and the low-frequency noise in the output leads. On the other hand, analytic results for the current fluctuations in Cooper pair splitters are scarce.  

In this work we investigate theoretically the current fluctuations in a Cooper pair splitter using techniques from full counting statistics.\cite{Bagrets2003,Belzig2002,Braggio2011,Flindt:Current,Flindt2005,Flindt2008,Flindt2010} In a recent article, some of us considered the distribution of waiting times between emitted electrons, and we showed that it contains a  wealth of information about the Cooper pair splitter, for instance the characteristic time scales that govern the underlying tunneling processes.\cite{Walldorf2018} Measurements of electron waiting times, however, require real-time detection of the individual tunneling events.\cite{gorman_tunneling_2017,jenei_2019,matsuo_2019} By contrast, conventional quantum transport experiments typically measure the electric currents and their fluctuations,\cite{Beckmann2004,Russo2005,Deacon:Cooper,Bruhat2018,Hofstetter:Cooper,Hofstetter2011,Das2012,Fulop2015,Ueda2019,Herrmann:Carbon,Herrmann:Spectroscopy,Schindele:Near,Fueloep:Local,Schindele:Nonlocal,Tan:Cooper,Borzenets:High} which are thus our main focus here. In particular, we consider the noise power spectra of the currents in the output leads\cite{Flindt2005c,Lambert:Nonequilibrium,Emary2007,Marcos2010,Ubbelohde2012} and their power-power correlations, which we use to analyze the physical processes involved in the splitting of Cooper pairs. We corroborate our findings by considering the $g^{(2)}$-function of the output currents,\cite{Emary2012,PhysRevB.99.085418} which provides an alternative view on the charge transport in the time domain. While earlier works have analyzed the shot noise of Cooper pair splitters using numerical approaches, we here employ projection operator methods that have been developed in the context of full counting statistics\cite{Flindt:Current,Flindt2005,Flindt2008,Flindt2010} and which allow us to carry out all calculations analytically. We are hereby able to develop a detailed understanding of the process of Cooper pair splitting and the associated fluctuations, which is relevant for future experiments on Cooper pair splitters. 

The rest of the paper is organized as follows. In~Sec.~\ref{Sec:Model}, we introduce the full Hamiltonian of the Cooper pair splitter, and we discuss how the combined system of a large-gap superconductor coupled to the quantum dots can be described by an effective Hamiltonian. In Sec.~\ref{Sec:Master}, we derive a quantum master equation for the dynamics of electrons in the quantum dots, which is valid with large negative biases on the drains. By dressing the quantum master equation with counting fields, we gain access to the full statistics of electrons that have tunneled into the drains and the corresponding current fluctuations. In~Sec.~\ref{Sec:Current}, we calculate the average currents and compare our results to earlier works before moving on to the noise power spectra of the output currents in~Sec.~\ref{Sec:Current-correlations}. Here, we first show how the zero-frequency noise allows for a simple and transparent interpretation of the charge transport in terms of contributions from elastic cotunneling between the quantum dots and the splitting of Cooper pairs, respectively. We then analyze the full frequency-dependent noise spectra and find that characteristic frequencies associated with Cooper pair splitting and elastic cotunneling, respectively, can be clearly identified in the finite-frequency noise spectra, thus providing experimental signatures of the two types of processes and their relative strengths. In Sec.~\ref{sec:power-power}, we consider higher-order cumulants of the currents. Our quantum master equation provides access to the full counting statistics of transferred electrons, and we here discuss the fourth cumulant of the currents, including the power-power correlations in the output leads. In~Sec.~\ref{Sec:SDC}, we turn to time-domain observables, and we show how our preceding analysis can be supported by investigations of the $g^{(2)}$-function of the output currents. Finally, in Sec.~\ref{Sec:Conclusion}, we give our conclusions, while technical details are provided in the Appendices.

\section{Cooper pair splitter}\label{Sec:Model}

Figure~\ref{fig:setup} shows the Cooper pair splitter consisting of two quantum dots in proximity to a superconductor that acts as a source of Cooper pairs. Strong Coulomb interactions on the quantum dots ensure that split Cooper pairs tunnel into different dots and further on into the separate normal-metal leads that act as electronic drains. The eigenstates of the uncoupled quantum dots are given by the occupation of each dot including the spin-degree of freedom. With a large superconducting gap, the proximity to the superconductor coherently couples the occupation states with the same particle parity, that is, an even or an odd number of particles. The even states with zero or two electrons are coupled by the process of Cooper pair splitting, where two electrons enter the quantum dots from the superconductor or vice versa.  The odd states with just a single electron on one of the dots are coupled by the process of elastic cotunneling, where an electron is transferred from one dot to the other via the superconductor. Under these conditions, the quantum dots and the superconductor can be described by an effective Hamiltonian as we discuss below.\cite{Sauret:Quantum,Konig} 

We start by specifying the full Hamiltonian 
\begin{equation}
\hat{H}=\hat{H}_{QD}+\hat{H}_{SC}+\hat{H}_N+\hat{H}_{T_S}+\hat{H}_{T_N},
\end{equation}
which describes the quantum dots, the superconductor, and the normal-metal leads, given by the first three terms, as well as the coupling between them given by the two tunneling Hamiltonians, $\hat{H}_{T_S}$ and $\hat{H}_{T_N}$, which we detail below. The Hamiltonian of the dots reads
\begin{equation}
\hat{H}_{QD}=\sum_{\ell\sigma}\epsilon_\ell^{\phantom\dagger}\hat{d}_{\ell\sigma}^\dagger \hat{d}_{\ell\sigma}^{\phantom\dagger}+\sum_{\ell} U_\ell \hat{n}_{\ell\uparrow}\hat{n}_{\ell\downarrow},\label{eq:HQD}
\end{equation}
where we have defined the operators  $\hat{d}_{\ell\sigma}^\dagger$ and $\hat{d}_{\ell\sigma}^{\phantom\dagger}$ that create and annihilate electrons with energy $\epsilon_{\ell}$ and spin~$\sigma$ in the left or right quantum dot, $\ell=L,R$. Here, the on-site interaction on the dots is denoted by $U_{\ell}$, and $\hat{n}_{\ell\sigma}\equiv \hat{d}_{\ell\sigma}^{\dagger}\hat{d}_{\ell\sigma}^{\phantom\dagger}$ counts electrons on the dots with spin $\sigma$. The superconductor is described by the BCS Hamiltonian
\begin{equation}
\hat{H}_{SC}=\sum_{q\sigma}\!\epsilon_{q}^{\phantom\dagger}\hat{a}_{q\sigma}^\dagger  \hat{a}_{q\sigma}^{\phantom\dagger}\!-\!\left(\sum_{q}\!\Delta^{\phantom\dagger}\hat{a}_{q\uparrow}^\dagger \hat{a}_{-q\downarrow}^\dagger\!+\!\text{h.c.}\right),
\end{equation}
where the operators $\hat{a}_{q \sigma}^\dagger$ and $\hat{a}_{q \sigma}^{\phantom\dagger}$ create and annihilate particles with momentum $q$ single-particle energy $\epsilon_{q}$ in the superconductor with the superconducting order parameter $\Delta$. The normal-state leads are described by the Hamiltonian
\begin{equation}
\hat{H}_N=\sum_{\ell k\sigma}\epsilon_{\ell k}^{\phantom\dagger}\hat{c}_{\ell k\sigma}^\dagger   \hat{c}_{\ell k\sigma}^{\phantom\dagger},\\
\end{equation}
while the coupling between the quantum dots and the external reservoirs are given by the tunneling Hamiltonians
\begin{equation}
\hat{H}_{T_S}=\sum_{\ell q \sigma}\left(t^{\phantom\dagger}_{S\ell q}\hat{a}^\dagger_{ q \sigma} \hat{d}^{\phantom\dagger}_{\ell \sigma}+\text{h.c.}\right)\label{eq:HTS}
\end{equation}
and
\begin{equation}
\hat{H}_{T_N}=\sum_{\ell k \sigma}\left(t^{\phantom\dagger}_{\ell k}\hat{c}^\dagger_{\ell k \sigma} \hat{d}^{\phantom\dagger}_{\ell \sigma}+\text{h.c.}\right),
\end{equation}
where $t_{S\ell q}$ and $t_{\ell k}$ are the tunneling amplitudes.

In the following, we consider strong Coulomb interactions on the quantum dots, such that each of them can be occupied by maximally one electron at a time. With a large superconducting gap, the combined system of the quantum dots and the superconductor can then be described by the effective Hamiltonian 
\cite{Sauret:Quantum,Konig}
\begin{equation}\label{eq:Heff}
\begin{split}
\hat{H}_S=&\sum_{\ell\sigma}\epsilon_\ell^{\phantom\dagger}\hat{d}_{\ell\sigma}^\dagger \hat{d}_{\ell\sigma}^{\phantom\dagger}-\hbar\gamma_\text{EC}\sum_\sigma\!\left(\hat{d}_{L\sigma}^\dagger \hat{d}_{R\sigma}^{\phantom\dagger}+\text{h.c.}\!\right)\\
&-\frac{\hbar\gamma_\text{CPS}}{\sqrt{2}}\!\left(\hat{d}_{L\downarrow}^\dagger \hat{d}_{R\uparrow}^\dagger-\hat{d}_{L\uparrow}^\dagger \hat{d}_{R\downarrow}^\dagger\!+\text{h.c.}\right)
\end{split}
\end{equation}
where $\hbar\gamma_\text{EC}$ and $\hbar\gamma_\text{CPS}$ are the amplitudes for elastic cotunneling and Cooper pair splitting. A detailed derivation of this Hamiltonian is provided in Appendix~\ref{AppA}.
 
In summary, we use the following operating conditions
\begin{equation}
k_BT, \epsilon_\ell, \hbar\gamma_{\mathrm{EC}}, \hbar \gamma_{\mathrm{CPS}}, \hbar\gamma_\ell\ll |e V_{\ell}|< \Delta <U_{\ell}, 
\end{equation}
where $V_{\ell}$ are the negative voltages applied to the drain electrodes, the temperature of the environment is denoted by $T$, and $\gamma_\ell$ are the tunneling rates from the quantum dots to the drains, which we introduce below. 
In this regime, we can trace out the normal-state electrodes and obtain a quantum master equation for the coupled quantum dots as shown in Appendix~\ref{AppDiss}.

\section{Quantum Master Equation}\label{Sec:Master}

Under the conditions specified above, the charge transport is unidirectional from the superconductor to the normal-state electrodes via the quantum dots. The system dynamics can then be described by a Markovian quantum master equation for the reduced density matrix~$\hat \rho$, defined in the Hilbert space of $\hat H_{S}$, reading\cite{Breuer}
\begin{equation}\label{eq:vonNeumannEq}
\frac{d}{dt}\hat \rho=\mathcal{L}\hat \rho=-\frac{i}{\hbar}[\hat H_{S},\hat \rho]+\mathcal{D}[\hat \rho].
\end{equation}
The Liouvillian $\mathcal{L}$ is the sum of the coherent evolution of the system itself, given by the commutator of the Hamiltonian $\hat H_S$ and the density matrix, and the dissipator
\begin{equation}\label{eq:Dissipator}
\mathcal{D}[\hat \rho]=\!\!\!\sum_{\sigma,\ell=L,R}\!\!\!\!\gamma_\ell^{\phantom\dagger}\left(\hat d_{\ell\sigma}^{\phantom\dagger}\hat \rho \hat d_{\ell\sigma}^\dagger-\frac{1}{2}\{\hat \rho,\hat d_{\ell\sigma}^\dagger \hat d_{\ell\sigma}^{\phantom\dagger}\}\right)\!,
\end{equation}
which describes the incoherent tunneling of electrons with spin~$\sigma=\uparrow,\downarrow$ from the left (right) quantum dot to the left (right) electrode at the rate $\gamma_\ell$, $\ell=L,R$. 

To evaluate the charge transport statistics, we resolve the density matrix with respect to the number of electrons that have tunneled into each of the normal-state leads during the time span $[0,t]$.\cite{Plenio1998,Makhlin2001} Thus, we introduce the $n$-resolved density matrix, $\hat \rho(\mathbf{n})$, where the vector $\mathbf{n}=(n_{L},n_{R})$ contains the number of transferred electrons. By tracing over the system degrees of freedom, we obtain the full counting statistics of transferred charge as
\begin{equation}
P(\mathbf{n},t)=\mathrm{Tr}[\hat \rho(\mathbf{n},t)].
\end{equation}
The unresolved density matrix is recovered as $\hat \rho(t)=\sum_\mathbf{n}\hat \rho(\mathbf{n},t)$. Moreover, it is convenient to introduce a vector of counting fields, $\boldsymbol{\chi}=(\chi_{L},\chi_{R})$, that couple to the number of transferred charges, by defining
\begin{equation}
\hat \rho(\boldsymbol{\chi},t)=\sum_\mathbf{n} \hat \rho(\mathbf{n},t)e^{i\mathbf{n}\cdot\boldsymbol{\chi}},
\end{equation}
whose equation of motion follows from Eq.~(\ref{eq:vonNeumannEq}) and reads
\begin{equation}\label{eq:vonNeumannEq_chi}
\begin{split}
\frac{d}{dt}\hat \rho(\boldsymbol{\chi},t)&=\mathcal{L}(\boldsymbol{\chi})\hat \rho(\boldsymbol{\chi},t)\\
&=\Big[\mathcal{L}+\!\!\!\sum_{\ell=L,R}\!\!\!(e^{i\chi_{\ell}}\!-\!1)\mathcal{J}_{\ell}\Big]\hat \rho(\boldsymbol{\chi},t).
\end{split}
\end{equation}
Here we have identified the jump operators that describe the transfer of an electron  into lead $\ell$ as
\begin{equation} 
\mathcal{J}_{\ell}\hat \rho=\gamma_\ell \sum_\sigma\hat d_{\ell\sigma}^{\phantom\dagger}\hat \rho \hat d_{\ell\sigma}^\dagger.
\label{eq:jumpop}
\end{equation}
Equation~(\ref{eq:vonNeumannEq_chi}) provides us with a complete description of the charge transfer statistics on all relevant time scales, and it allows us to evaluate quantities such as the distribution of electron waiting times,\cite{Brandes:Waiting,Thomas2013,Walldorf2018} the noise power spectra of the currents,\cite{Flindt2005c,Lambert:Nonequilibrium,Emary2007,Marcos2010,Ubbelohde2012} and the full counting statistics of the transferred charge.\cite{Flindt:Current,Flindt2005,Flindt2008,Flindt2010} In the following sections, we use Eq.~(\ref{eq:vonNeumannEq_chi}) to investigate the current fluctuations in the Cooper pair splitter.
 
\section{Average current}\label{Sec:Current}

We start by considering the mean current flowing from the superconductor into the drain electrodes. Throughout this work, we consider particle currents instead of electrical currents, since it allows us to omit powers of the electron charge. Due to charge conservation, the current from the superconductor can be written as
\begin{equation}
\langle I_S \rangle= \langle I_L \rangle+\langle I_R\rangle
\end{equation}
in terms of the currents running into the normal-state drains, $\langle I_\ell\rangle$, $\ell=L,R$, which can be expressed as
\begin{equation}
\langle I_\ell\rangle = \text{Tr}[\mathcal{J}_\ell\hat\rho_S],
\end{equation}
where the stationary state $\hat\rho_S$ is given by the normalized solution to $\mathcal{L}\hat \rho_S=0$. The current from the superconductor then becomes
\begin{equation}
\label{eq:av_curr}
\langle I_S \rangle=\bar{\gamma}_\text{CPS}^2\gamma_\Sigma,
\end{equation}
where we have introduced the average rate
\begin{equation}
\gamma_\Sigma=(\gamma_L+\gamma_R)/2,
\end{equation}  
and defined the renormalized couplings
\begin{equation}
\bar{\gamma}_\text{CPS}^2 = \frac{4\gamma_\text{CPS}^2}{\varepsilon^2+\gamma_\Sigma^2+4\gamma_\text{CPS}^2/\eta},
\end{equation}
and
\begin{equation}
\bar{\gamma}_\text{EC}^2=\frac{4\gamma_\text{EC}^2}{\delta^2+\gamma_\Sigma^2+4\gamma_\text{EC}^2},
\end{equation}
where $\delta=(\epsilon_L-\epsilon_R)/\hbar$ and $\varepsilon=(\epsilon_L+\epsilon_R)/\hbar$ are the detuning and the sum of the energy levels, respectively.  In addition, we have introduced the parameter
\begin{equation}
\eta=1+\left(\frac{\gamma_L-\gamma_R}{2\gamma_\Sigma}\right)^2\left[(\bar{\gamma}_\text{EC})^2-1\right],
\end{equation}
which reduces to one for a symmetric setup with~$\gamma_L=\gamma_R$. We note that the expression for the current recovers the result of  Ref.~\onlinecite{Sauret:Quantum} obtained with $\gamma_\text{EC}=0$ and the  energy renormalization absorbed into the dot levels as discussed at the end of Appendix~\ref{AppA}. In addition, for $\gamma_\text{CPS}\ll\gamma_\Sigma$, we reproduce the result of Ref.~\onlinecite{Recher:Andreev} in that limit.

The average current is shown in Fig.~\ref{fig:setup}b as a function of the quantum dot energies. The current is maximal along the line $\varepsilon=0$, where the doubly occupied state is on resonance with the empty state, and Cooper pair splitting is energetically favorable. Along this resonance line, the current is only weakly dependent on the detuning of the energy levels, $\delta$, as shown in Fig.~\ref{fig:setup}c. Moreover, for a symmetric setup with $\gamma_L=\gamma_R$, the elastic cotunneling processes do not influence the average current, which becomes independent of the detuning (not shown). In Fig.~\ref{fig:setup}c, we also show the average current away from the resonance condition $\varepsilon=0$, and the process of Cooper pair splitting gets suppressed. The peak in the current is Lorentzian with a broadening given by the coupling to the external electrodes. We also note that the elastic cotunneling processes are enhanced, when the quantum dot levels are on resonance, meaning that the detuning vanishes, $\delta=0$.

\section{Noise power spectrum}\label{Sec:Current-correlations}

\begin{figure*}
  \centering
  \includegraphics[width=1\textwidth]{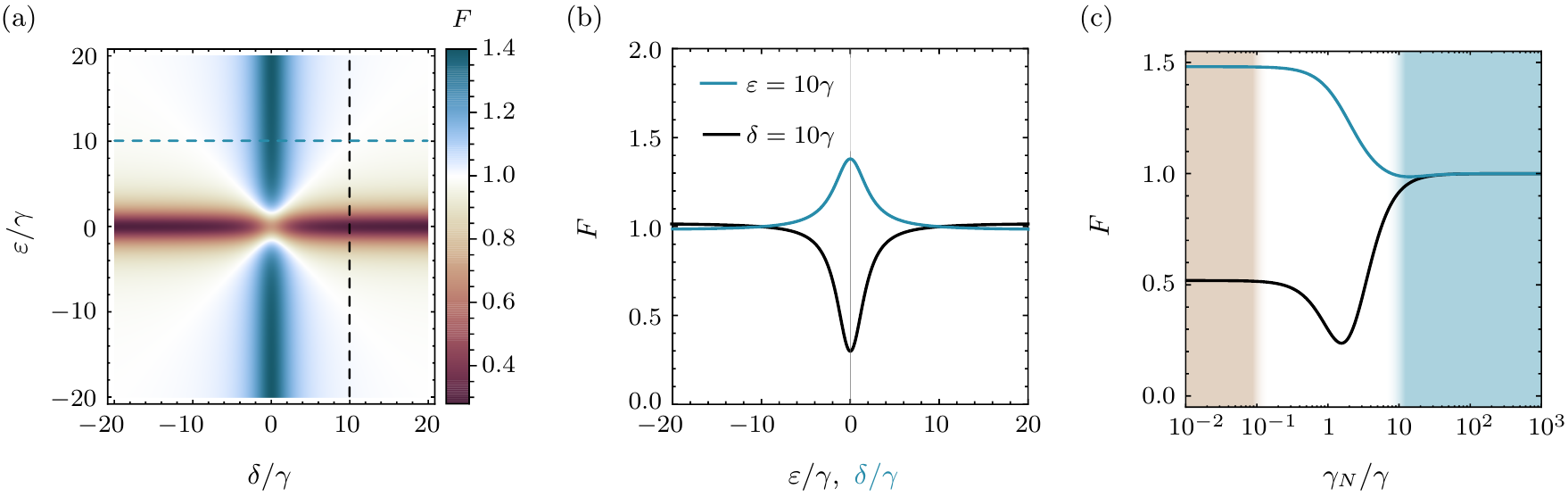}
  \caption{Fano factor of the autocorrelations. (a) Fano factor as a function of the detuning, $\delta$, and the total energy, $\varepsilon$, of the quantum dot levels with  $\gamma_L=\gamma_R=\gamma_\text{CPS}=\gamma_\text{EC}\equiv\gamma$. (b) Fano factor along the cuts indicated in the left panel. (c)~Fano factor as a function of $\gamma_L=\gamma_R\equiv\gamma_N$ with $\gamma_\text{CPS}=\gamma_\text{EC}\equiv\gamma$ and $\epsilon_L=-\epsilon_R=5\hbar\gamma$ (black line) and $\epsilon_L=\epsilon_R=5\hbar\gamma$ (blue line).}
\label{fig:Fano}
\end{figure*}

We next investigate the fluctuations of the current. To this end, we consider the noise power spectrum of the tunnel currents between the quantum dots and the drains. The noise power spectrum reads\cite{BLANTER20001}
\begin{equation}\label{eq:Noise}
S_{\ell\ell'}(\omega)=\frac{1}{2}\int_{-\infty}^{\infty}\text{d}te^{i\omega t}\langle\{\delta \hat{I}_{\ell}(t),\delta \hat{I}_{\ell'}(0)\}\rangle,
\end{equation}
where the operator $\delta \hat{I}_{\ell}(t)=\hat{I}_{\ell}(t)-\langle \hat{I}_{\ell}(t)\rangle$, $\ell=L,R$, measures the deviation of the tunnel current from its average value, and curly brackets denote an anti-commutator. The autocorrelation spectrum, $S_{\ell\ell}(\omega)$, is always real and positive. By contrast, the cross correlations, $S_{\ell\neq\ell'}(\omega)$, can take complex values at finite frequencies, but we only consider the real part, and from now on we let $S_{\ell\neq\ell'}(\omega)$ denote the real part. Below, we do not need to specify the current operators. Instead, MacDonald's formula\cite{MacDonald1962} allows us to relate the noise power spectrum to the quantum master equation~(\ref{eq:vonNeumannEq_chi}) and express it as~\cite{Flindt2008,Flindt2005c,Brandes:Waiting} 
\begin{equation}\label{eq:Macdonald}
S_{\ell\ell'}(\omega)=\delta_{\ell\ell'}\text{Tr}[\mathcal{J}_\ell\hat\rho_S]-\text{Re}\!\left\{\text{Tr}\left[\mathcal{J}_\ell \mathcal{R}(\omega)\mathcal{J}_{\ell'}\hat\rho_S\right]\!+\!(\ell\!\leftrightarrow\!\ell')\right\}\!,
\end{equation} 
where the pseudoinverse, $\mathcal{R}(\omega)$, is defined as~\cite{Flindt2005,Flindt2005c,Flindt2008,Flindt2010} 
\begin{equation}
\mathcal{R}(\omega)=\mathcal{Q}(\mathcal{L}+i\omega)^{-1}\mathcal{Q},
\label{eq_pseudo} 
\end{equation}
in terms of the orthogonal projectors $\mathcal{Q}=1-\mathcal{P}$ and $\mathcal{P}[\ \cdot\ ]=\hat{\rho}_S\mathrm{Tr}[\ \cdot \ ]$. The pseudoinverse is well-defined even for $\omega=0$, since the inversion is performed only in the subspace spanned by $\mathcal{Q}=1-\mathcal{P}$, where $\mathcal{L}$ is regular, since the null space has been projected away. Using the matrix representation of the Liouvillian in Appendix~\ref{AppC}, we can then evaluate the noise spectrum. Details on how to evaluate the pseudoinverse can be found on page 7 of Ref.~\onlinecite{Flindt2010}. 

\begin{widetext}
Interestingly, the noise power spectrum can be determined analytically. Specifically, for a symmetric setup with $\gamma_L=\gamma_R=\gamma_N$,  we find for the Fano factor, $F_{{\ell\ell'}}(\omega)=S_{\ell\ell'}(\omega)/I_N$, the expression
\begin{equation}
F_{{\ell\ell'}}(\omega) = \delta_{{\ell\ell'}}- I_N\gamma_N(\gamma_N^2+\omega_\text{CPS}^2)\Bigg(\frac{5\gamma_N^2+\omega_\text{CPS}^2+\omega^2}{h(\omega_\text{CPS},\omega)}-\frac{(1-\delta_{{\ell\ell'}})}{2\gamma_\text{CPS}^2(\gamma_N^2+\omega^2)}+(-1)^{\delta_{{\ell\ell'}}}\!\left[ \frac{\gamma_\text{EC}}{\gamma_\text{CPS}}\right]^2\!\frac{\gamma_N^2+\omega_\text{EC}^2-3\omega^2}{h(\omega_\text{EC},\omega)}\Bigg),
\label{Correlation spectra}
\end{equation}
having defined the average current $I_N \equiv \langle I_L\rangle=\langle I_R\rangle=\langle I_S\rangle/2$ and the characteristic frequencies $\omega_\text{CPS} = \sqrt{4\gamma_\text{CPS}^2+\varepsilon^2}$ and 
$\omega_\text{EC} = \sqrt{4\gamma_\text{EC}^2+\delta^2}$, as well as the function 
\begin{equation}
h(\omega_0,\omega)= (\gamma_N^2+\omega^2)^3+2(\gamma_N^4-\omega^4)\omega_0^2+(\gamma_N^2+\omega^2)\omega_0^4.
\end{equation}
\end{widetext}
With this expression in hand, we now discuss the information about the charge transport that we can extract. 

To begin with, we consider the zero-frequency component of the current correlations. The zero-frequency noise of a Cooper pair splitter has previously been calculated numerically in Ref.~\onlinecite{Sauret:Spin}. Here, we obtain a compact expression for the zero-frequency noise reading
\begin{equation}
F_{{\ell\ell'}}=1+\left(\delta_{\ell\ell'}-\frac{1}{2}\right)\bar{\gamma}_\text{EC}^2-\frac{I_N}{\gamma_N}\left(1+\frac{2I_N\gamma_N}{\gamma^2_\text{CPS}}\right),
\label{eq:noisezf}
\end{equation}
which provides an interesting interpretation of the charge transport. In the absence of elastic cotunneling and with large tunneling rates to the drains, only the first term survives, and the Fano factors equal unity. In this case, the separate flows of electrons into each drain resemble a Poisson process. However, the currents are correlated, since electrons are injected pairwise from the superconductor as split Cooper pairs. For this reason, the Fano factor of the cross-correlations is positive and not zero as one would expect for two uncorrelated processes. 

\begin{figure*}
	\centering
	\includegraphics[width=1\textwidth]{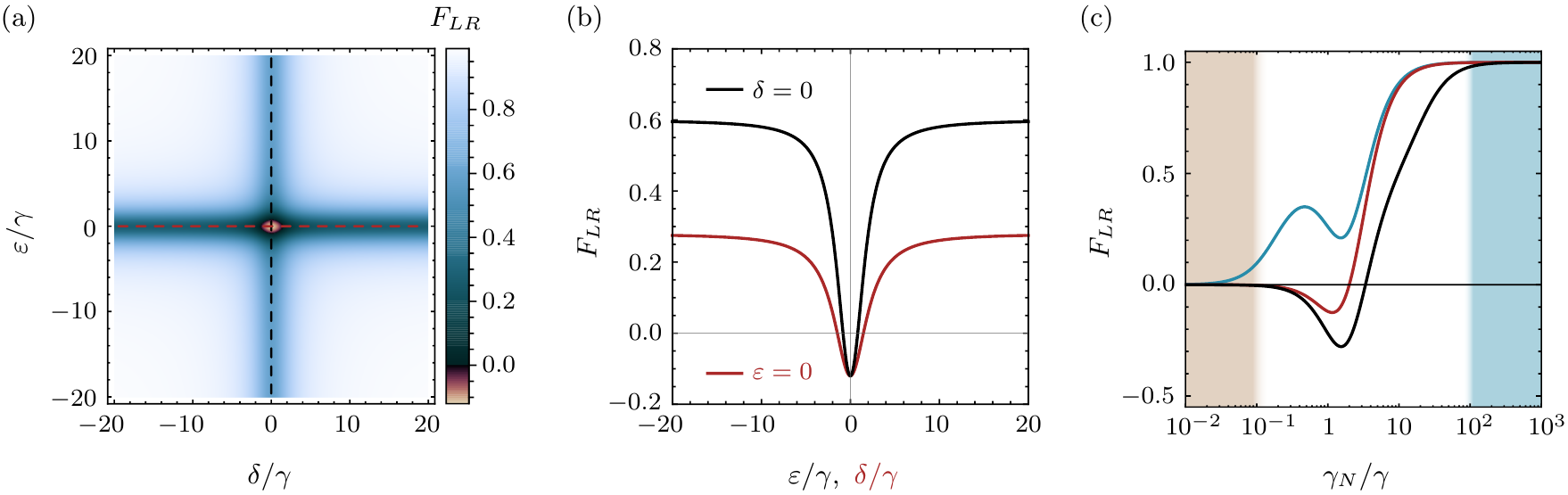}
	\caption{Fano factor of the cross-correlations. (a) Fano factor as a function of the detuning, $\delta$, and the total energy, $\varepsilon$, of the quantum dot levels with  $\gamma_L=\gamma_R=\gamma_\text{CPS}=\gamma_\text{EC}\equiv\gamma$. (b) Fano factor along the cuts indicated in the left panel. (c) Fano factor as a function of $\gamma_L=\gamma_R\equiv\gamma_N$ with $\epsilon_L=\epsilon_R=0$, $\gamma_\text{CPS}\equiv\gamma$, and  $\gamma_\text{EC}=0.1\gamma$ (blue line), $1\gamma$ (red line), $10\gamma$ (black line).}
	\label{fig:cross}
\end{figure*}

This picture gradually breaks down as elastic cotunneling is included, and the second term becomes non-zero. Elastic cotunneling reduces the  correlations between the currents in the drains, since electrons are allowed to transfer between the quantum dots. In this way, the separate flows of electrons into the drains get mixed, which both reduces their correlations and increases the fluctuations in each drain. Furthermore, as the coupling to the drain electrodes is lowered, also the third term becomes important, and it reduces both the auto and the cross correlations. In this case, the lowered coupling to the leads introduces a finite dwell time of electrons on the quantum dots, which reduces the fluctuations in the leads and the correlations between the currents.

To gain further insight into the current fluctuations, we show in Figs.~\ref{fig:Fano} and~\ref{fig:cross} the Fano factors of the auto and cross correlations, respectively. In Fig.~\ref{fig:Fano}a, the splitting of Cooper pairs is favorable along the resonance line, $\varepsilon=0$, and the Fano factor is suppressed well below one due to the tunnel barriers between the quantum dots and the drains. By contrast, along the other resonance line, $\delta=0$, where elastic cotunneling is enhanced, the fluctuations are increased, since the separate flows of electrons get mixed. These effects are also illustrated in Fig.~\ref{fig:Fano}b, where we show the Fano factor along the cuts in the left panel, which both cross one of the resonance lines. In Fig.~\ref{fig:Fano}c, we show the Fano factor as a function of the coupling to the drain electrodes. In the blue-shaded region, electrons immediately leave the quantum dots via the drains because of the large coupling, and the Fano factor approaches unity, signaling that the injection of split Cooper pairs becomes a Poisson process. In the brown-shaded region, the coupling is very low, and the Fano factor now depends strongly on the energy levels of the quantum dots. For $\varepsilon=0$ (black line), Cooper pair splitting is favorable, and the occupations of the quantum dots oscillate between being empty and doubly-occupied. In that case, the quantums dots are occupied half of the time, and the Fano factor is suppressed accordingly. For $\delta=0$ (blue line), elastic cotunneling is enhanced, and the fluctuations are increased due to the mixing of the separate flows of electrons. In between these parameter regimes, the Fano factor develops a more complicated structure, since all possible processes are combined.

In Fig.~\ref{fig:cross}, we turn to the Fano factor of the cross-correlations. In Fig.~\ref{fig:cross}a, we observe a large degree of correlation \emph{away} from the resonance lines. In that case, neither Cooper pair splitting nor elastic cotunneling are favorable. Still, once a split Cooper pair is injected into the quantum dots and one electron tunnels out via a drain electrode, the other electron likely leaves via the other drain electrode, leading to the large correlations. However, despite the large correlations, the actual currents are of course small, since the system is operated away from any of the important resonance conditions. In Fig.~\ref{fig:cross}b, we consider the cross-correlations along the two resonance lines, $\delta=0$ (black line) and $\varepsilon=0$ (red line), where either elastic cotunneling or Cooper pair splitting is favorable. Elastic cotunneling reduces the cross-correlations, since it mixes the separate flows of electrons. They also get reduced, if Cooper pair splitting is on resonance, and electrons quickly oscillate back and forth between the superconductor and the quantum dots. When the two processes are combined, we even observe negative cross-correlations between the output currents as seen in the figure.

Finally, in Fig.~\ref{fig:cross}c, we consider the cross-correlations as a function of the coupling to the drain electrodes, and again we can identify three distinct regimes. For low couplings in the brown-shaded region, the tunneling events into the drains are rare and uncorrelated. By contrast, in the blue-shaded region, where the coupling is large, split Cooper pairs are immediately evacuated from the quantum dots via the drains, leading to strong correlations. In between these parameter regimes, the cross-correlations are more complicated as discussed above.

\begin{figure*}
  \centering
  \includegraphics[width=1\textwidth]{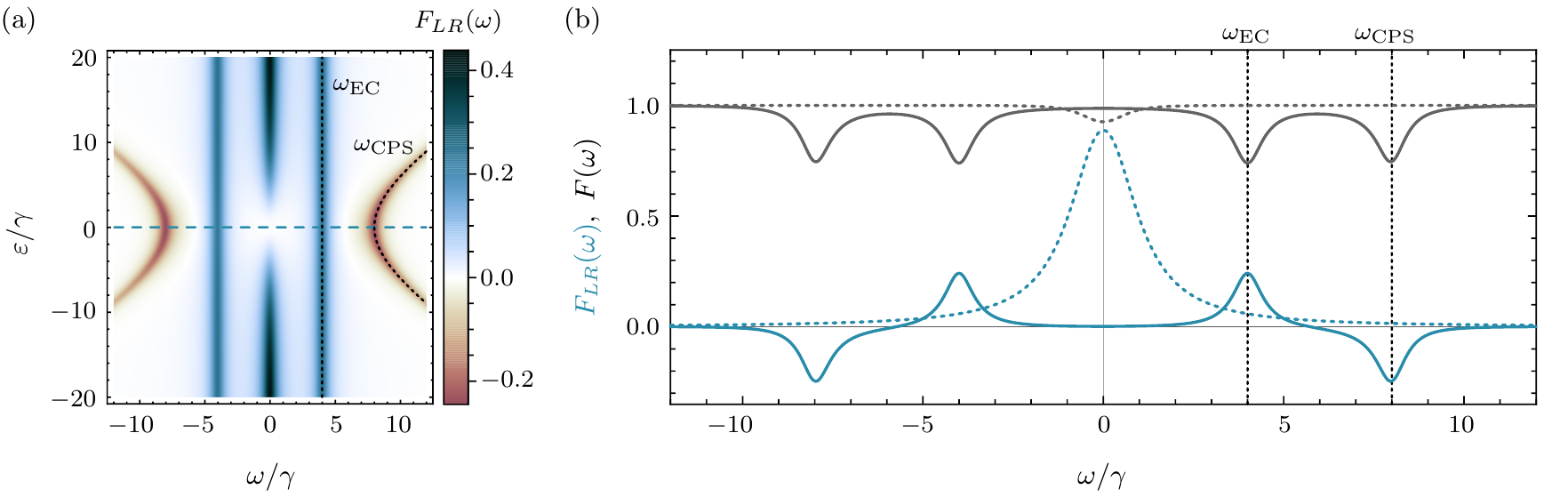}
  \caption{Finite-frequency noise. (a) Fano factor of the cross-correlations as a function of the frequency and the total energy of the quantum dots. The other parameters are $\gamma_L=\gamma_R\equiv0.5\gamma$, $\gamma_\text{CPS}=4\gamma$, $\gamma_\text{EC}=2\gamma$, and $\delta=0$. (b) Auto (gray) and cross (blue) correlations as functions of the frequency for $\epsilon_L=\epsilon_R=0$ and $\gamma_L=\gamma_R\equiv0.5 \gamma$, $\gamma_\text{CPS}=4\gamma$, $\gamma_\text{EC}=2\gamma$ (solid), given by the cut in the left panel, and $\gamma_L=\gamma_R\equiv\gamma$,  $\gamma_\text{CPS}=\gamma_\text{EC}=0.1 \gamma$ (dotted). In both panels, we indicate $\omega_{\mathrm{CPS}}$ and $\omega_{\mathrm{EC}}$. }
\label{fig:finite-freq}
\end{figure*}

Next, we consider the full frequency-dependent noise spectra given by Eq.~(\ref{Correlation spectra}) and displayed in Fig.~\ref{fig:finite-freq}. In Fig.~\ref{fig:finite-freq}a, we show the Fano factor of the cross-correlations as a function of the observation frequency and the total energy of the quantum dots. Of particular interest are the dips and peaks in the cross-correlations that appear at the characteristic frequencies, $\omega_\text{CPS}$ and $\omega_\text{EC}$, associated with Cooper pair splitting and elastic cotunneling, respectively, thus providing a direct experimental method to distinguish the two types of processes. The figure also illustrates how $\omega_\text{CPS}$ depends on the total energy, while $\omega_\text{EC}$ remains constant. In Fig.~\ref{fig:finite-freq}b, we show both the auto  and cross correlation spectra along the resonance line indicated in the left panel, and here we again see how the cross-correlations allow us to distinguish Cooper pair splitting from elastic cotunneling. By contrast, the two types of processes both lead to dips in the autocorrelation spectrum. We also see how a large coupling to the drain electrodes washes out these features, which might also not be robust against external decoherence and dephasing mechanisms that are not included here.

\section{Power-power correlations}
\label{sec:power-power}

Until now, we have focused on the average current and the noise power spectra, which at zero frequency correspond to the first and second cumulants of the currents. However, with the counting fields included in our quantum master equation, we can in principle access any cumulant of the full counting statistics. To this end, we formally solve the quantum master equation~(\ref{eq:vonNeumannEq_chi}) as $\hat \rho(\boldsymbol{\chi},t)=e^{\mathcal{L}(\boldsymbol{\chi})t}\hat \rho_S$,
assuming that the system has reached its stationary state at the time $t=0$, when the counting of particles begins. We also define the cumulant generating function for the charge transfer statistics as
\begin{equation} \label{eq:CGF}
S(\boldsymbol{\chi},t)=\ln\left[\sum_\mathbf{n} P(\mathbf{n},t)e^{i\mathbf{n}\cdot\boldsymbol{\chi}}\right]
=\ln\mathrm{Tr}\left[e^{\mathcal{L}(\boldsymbol{\chi})t}\hat \rho_S\right].
\end{equation}
We then see that the scaled cumulant generating function
\begin{equation}
\Theta(\boldsymbol{\chi})=\lim_{t\rightarrow\infty}\frac{S(\boldsymbol{\chi},t)}{t}=\max_i\{\lambda_{i}(\boldsymbol{\chi})\},
\label{eq:CGF}
\end{equation}
for long observation times is given by the eigenvalue of $\mathcal{L}(\boldsymbol{\chi})$ with the largest real part. For small values of the counting fields, this is the eigenvalue that develops adiabatically from the zero-eigenvalue corresponding to the stationary state. All other eigenvalues have negative real parts, causing the system to relax to its stationary state. 

All zero-frequency cumulants of the (particle) current can now be obtained by differentiating the cumulant generating function with respect to the counting fields as
\begin{equation}
\langle\!\langle I^n_\ell I^m_{\ell'}\rangle\!\rangle=\partial^n_{\chi_\ell}\partial^m_{\chi_{\ell'}}\Theta(\boldsymbol{\chi})|_{\boldsymbol{\chi}= \mathbf{0}},
\end{equation}
where double brackets denote cumulant averages. The first and second cumulants are the average currents and the zero-frequency noise, respectively. Here, we focus on the power-power correlations in the drains, $\langle\!\langle I^2_\ell I^2_{\ell'}\rangle\!\rangle$, i.e., the correlations between the squared currents in the output leads. Such correlations have not received much attention in the past, but they can in principle be measured, and they can be evaluated using our quantum master equation dressed with counting fields. Technically, we have to evaluate the derivatives of the eigenvalue of $\mathcal{L}(\boldsymbol{\chi})$ with the largest real part according to Eq.~(\ref{eq:CGF}). However, due to the large matrix size of $\mathcal{L}(\boldsymbol{\chi})$, we cannot directly evaluate its eigenvalues as functions of the counting fields. Instead, we find the derivatives of the largest eigenvalue using perturbation theory in the counting fields as discussed in Refs.~\onlinecite{Flindt2005,Flindt2008,Flindt2010}. The method takes the zero-eigenvalue and the stationary state as the starting point and then calculates corrections to the eigenvalue order-by-order in the counting fields to obtain cumulants of any order. The details of this perturbation scheme are outlined in Appendix~\ref{AppD}, and below we just quote the final results.

\begin{figure*}
  \centering
  \includegraphics[width=1\textwidth]{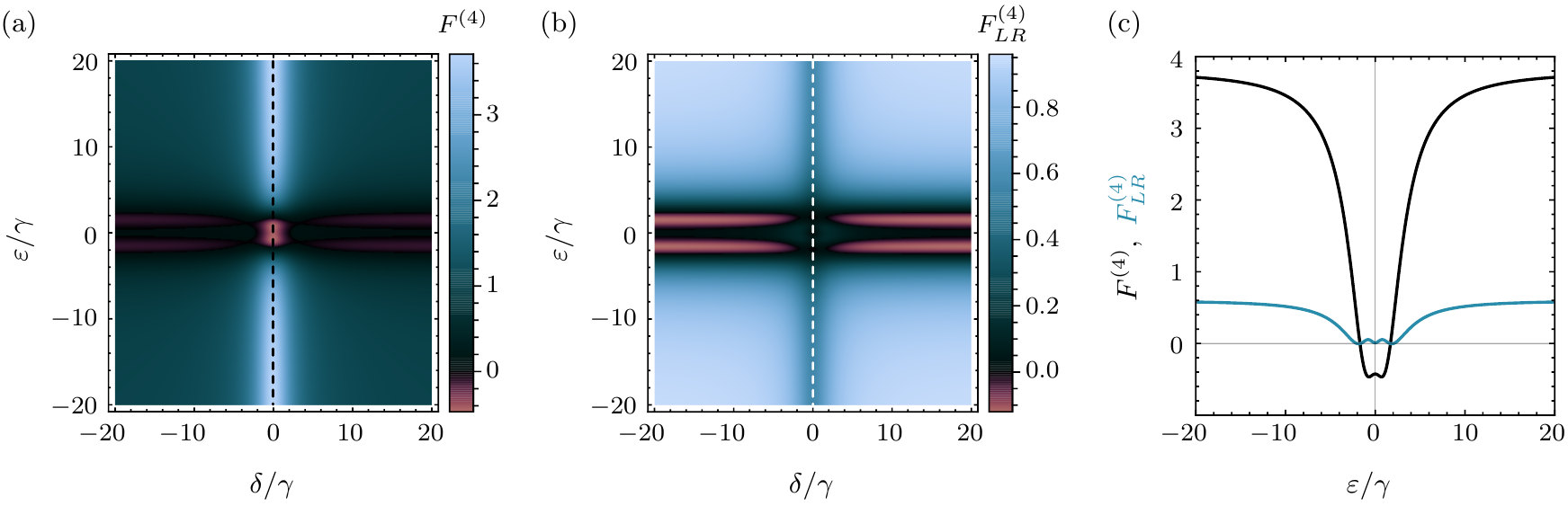}
  \caption{Power-power correlations. (a) Fano factor of the power-power autocorrelations as a function of the detuning, $\delta$, and the total energy, $\varepsilon$, of the quantum dot levels with $\gamma_L=\gamma_R=\gamma_\text{CPS}=\gamma_\text{EC}\equiv\gamma$. (b) Fano factor of the power-power cross-correlations for the same parameters. (c) Auto and cross correlations along the cuts indicated in the left panels, where $\delta=0$.}
\label{fig:power-power}
\end{figure*}

\begin{widetext}
For the autocorrelations of the power (or the fourth cumulant of the currents), we find 
\begin{equation}
\langle \! \langle I_\ell^4\rangle\!\rangle =\langle \! \langle I_\ell^2\rangle\!\rangle-12\text{Tr}\big[  \mathcal{J}_\ell\mathcal{R}\big\{1 +2 \mathcal{I}_{\ell} (1+\mathcal{R}\mathcal{I}_{\ell})\mathcal{R}+\mathcal S_{\ell}\mathcal{R}\big\}\mathcal J_\ell \hat\rho_S\big],
\label{Power-power auto}
\end{equation}
where $\mathcal{I}_{\ell} = \langle I_{\ell}\rangle-\mathcal J_{\ell}$ and $\mathcal S_{\ell} = \langle \! \langle  I^2_{\ell}\rangle\!\rangle-\mathcal{J}_\ell$ in terms of the zero-frequency noise, $\langle \! \langle  I^2_{\ell}\rangle\!\rangle=S_{\ell\ell}(0)$, and $\mathcal{R} = \mathcal{R}(0)$ is the pseudo-inverse in Eq.~(\ref{eq_pseudo}) evaluated at $\omega=0$. (We note that the perturbation scheme also yields the noise power spectrum in Eq.~(\ref{eq:Macdonald}) for $\omega=0$.) For the power-power correlations, we arrive at the more complicated expression
\begin{equation}
\langle \! \langle I_\ell^2 I_{\ell'}^2\rangle\!\rangle =\text{Tr}\big[ \mathcal{J}_\ell\mathcal{R} \big\{(1+2\mathcal{I}_{\ell} \mathcal{R})(1+2\mathcal I_{\ell'}\mathcal{R})\mathcal I_{\ell'} +4\mathcal I_{\ell'}\mathcal{R}(\mathcal I_{\ell'}\mathcal{R}\mathcal I_\ell+\mathcal I_\ell\mathcal{R}\mathcal I_{\ell'})+4\langle \! \langle I_\ell I_{\ell'}\rangle\!\rangle\mathcal{R} \mathcal I_{\ell'} +2\mathcal S_{\ell'}\mathcal{R}\mathcal I_\ell\big\}\hat\rho_S\big]+(\ell\leftrightarrow {\ell'}).
\label{Power-power correlation}
\end{equation}
We can now evaluate these formulas based on the Liouvillian $\mathcal{L}$ and the jump operators in Eq.~(\ref{eq:jumpop}). The resulting expressions are lengthy, and here we only present analytical results in certain limits together with figures.
\end{widetext}

For a symmetric Cooper pair splitter, where the amplitude for Cooper pair splitting is much smaller than the total energy of the quantum dots, the average current is suppressed, and the Fano factor, $ F^{(4)}_{\ell\ell'}=\langle \! \langle I_\ell^2 I_{\ell'}^2\rangle\!\rangle/I_N$ for the power-power correlations simplify to the expression
\begin{equation}
F^{(4)}_{\ell\ell'} = 1+\left(4\delta_{\ell\ell'}-\frac{1}{2}\right)\bar{\gamma}_\text{EC}^2+\mathcal{O}\left(\frac{\gamma_\text{CPS}}{\sqrt{\gamma_N^2+\varepsilon^2}}\right),
\end{equation}
where the higher-order terms are different for the auto and the cross correlation and depend on all parameters. Just as for the current-current correlations in Eq.~(\ref{eq:noisezf}) in that limit, we see that the autocorrelations are Poissonian, if elastic cotunneling is negligible. At the same time, the cross-correlations remain positive, since the two separate flows of electrons originate from the same random splitting of Cooper pairs. In this context, elastic cotunneling reduces the cross-correlations by mixing the two flows, and it also strongly increases the autocorrelations of the power fluctuations. More generally, we find that $F^{(4)}_{\ell\ell}=F^{(4)}_{\ell\ell'}$, if $\gamma_\text{EC} \ll \sqrt{\gamma_N^2+\delta^2}$, such that cotunneling is negligible.

In Fig.~\ref{fig:power-power}, we show the Fano factors of the auto and cross correlations of the power fluctuations as functions of the detuning and the total energy of the quantum dot levels. The fluctuations in each lead are generally large as we move along the resonance line, $\delta=0$, where elastic cotunneling is favorable. However, the fourth cumulant of the current gets reduced, and even becomes negative, as also Cooper pair splitting comes into resonance. The cross-correlations also get reduced, even if the average current is large on resonance, since the elastic cotunneling processes mix the separate flows of electrons and thereby destroy the correlations.

\section{Time-domain observables}\label{Sec:SDC}

\begin{figure*}
  \centering
  \includegraphics[width=1\textwidth]{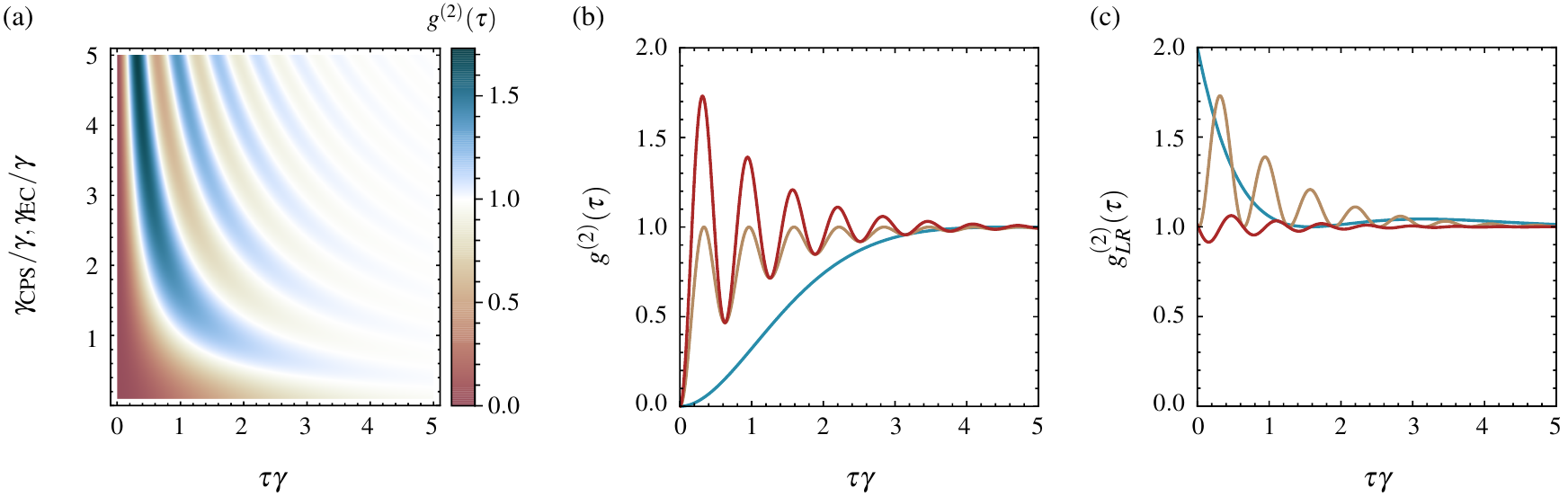}
  \caption{Time-domain observables. (a) The $g^{(2)}$-correlation function of the output currents as a function of the time and the amplitudes for Cooper pair splitting and elastic cotunneling being equal, $\gamma_\text{CPS}=\gamma_\text{EC}$. The other parameters are $\delta = \varepsilon =0$ and $\gamma_N=\gamma$. (b) Autocorrelation function, $g^{(2)}(\tau)$, for $\gamma_\text{CPS}=0.5\gamma$ and  $\gamma_\text{EC}=0$ (blue),  $\gamma_\text{CPS}=5\gamma$ and  $\gamma_\text{EC}=0$ (yellow), and $\gamma_\text{CPS}=5\gamma$ and  $\gamma_\text{EC}=5\gamma$ (red). (c) Cross-correlation function, $g^{(2)}_{LR}(\tau)$, for the same parameters as in the middle panel. }
\label{fig:g2}
\end{figure*}

Having investigated the current fluctuations in the frequency domain, we now change perspective and instead analyze the charge transport statistics in the time domain. In a recent work, we considered the distribution of waiting times between tunneling events into the drains.\cite{Walldorf2018} As an alternative, we here consider the $g^{(2)}$-function of the output currents. Based on our quantum master equation, the $g^{(2)}$-function can be obtained as\cite{Emary2012}
\begin{equation}
g^{(2)}_{\ell\ell'}(\tau) = \frac{\text{Tr}\left[\mathcal{J}_\ell e^{{\mathcal{L}}\tau}\mathcal{J}_{\ell'}\hat\rho_S\right]+\text{Tr}\left[ \mathcal{J}_{\ell'}e^{{\mathcal{L}}\tau}\mathcal{J}_\ell \hat \rho_S\right]}{2\langle I_\ell\rangle \langle I_{\ell'}\rangle},
\end{equation} 
where $\tau$ is the time between tunneling events described by the jump operators $\mathcal{J}_\ell$ and $\mathcal{J}_{\ell'}$. Here we consider a symmetrized $g^{(2)}$-function, although this makes no difference for the symmetric setup we consider below. The $g^{(2)}$-function is the probability that an electron tunnels into lead $\ell$ (or $\ell'$) at the time $\tau$ after an electron has tunneled into lead  $\ell'$ (or $\ell$), normalized with respect to the unconditional probability. Evaluating this expression for a symmetric setup, we find for the $g^{(2)}$-functions
\begin{widetext}
\begin{equation}
g^{(2)}_{\ell \ell'}(\tau) = 1-e^{-\gamma_N \tau}\Bigg[\!\!\left(\!\cos\!\left[\frac{\omega_\text{CPS}\tau}{2} \right] \!+\!\frac{\gamma_N}{\omega_\text{CPS}}\sin\!\left[\frac{\omega_\text{CPS}\tau}{2} \right]\right)^2
-g_x(1-\delta_{\ell \ell'})+(-1)^{\delta_{\ell \ell'}}g_x\left(\frac{2\gamma_\text{EC}\sin\left[\frac{\omega_\text{EC}\tau}{2} \right]}{\omega_\text{EC}}\right)^2\Bigg],
\label{g2 function}
\end{equation}
where we have defined the parameter $g_x =(\gamma_N^2+\omega_\text{CPS}^2)/(2\gamma_\text{CPS})^2 $.
\end{widetext}

We start by analyzing the $g^{(2)}$-function of the individual currents. Here, we first notice that $g^{(2)}_{ll}(0)=0$, which is a direct manifestation of the strong Coulomb interactions that prevent two electrons from being emitted from the same dot simultaneously. Furthermore, we find that $g^{(2)}_{ll}(0) < g^{(2)}_{ll}(\tau)$ for $\tau>0$, implying that the electron emission from each quantum dot is always anti-bunched, even if the emission statistics may be super-Poissonian.\cite{Emary2012} 

In the first two panels of Fig.~\ref{fig:g2}, we show the $g^{(2)}$-function of the individual currents and observe an oscillatory pattern that is washed out as the coupling to the drain electrodes is increased. In particular, if the coupling is much larger than the characteristic frequency associated with Cooper pair splitting, $\gamma_N \gg \omega_\text{CPS}$, and the frequency associated with elastic cotunneling is small, $\omega_\text{EC} \simeq 0$, we find
\begin{equation} 
g_{\ell\ell}^{(2)}(\tau) \simeq 1-e^{-\gamma_N \tau}\left(1+\frac{\gamma_N \tau}{2}\right)^2,
\end{equation} 
which increases monotonously with time. In the other extreme, where the coupling is smaller than the frequency of Cooper pair splitting, $\omega_\text{CPS} \gtrsim \gamma_N$, an oscillatory pattern with frequency $\omega_\text{CPS}$ appears due to the coherent oscillations between the quantum dots and the superconductor. Similarly, for $\omega_\text{EC} \gtrsim \gamma_N$, elastic cotunneling leads to oscillations, however, with frequency $\omega_\text{EC}$.

In Fig.~\ref{fig:g2}c, we turn to the $g^{(2)}$-function of the cross-correlations. In this case, we find at short times
\begin{equation}
g_{\ell\neq\ell'}^{(2)}(0) = 1 +\frac{\gamma_N^2+\omega_\text{CPS}^2}{4\gamma_\text{CPS}^2},
\end{equation}
showing that the probability for simultaneous emissions into the left and right drain electrodes increases with the coupling to the leads, $\gamma_N$, and the total energy, $|\varepsilon|$. By contrast, as one might expect, elastic cotunneling has no effect on $g_{\ell\ell'}^{(2)}(\tau)$ on short timescales, $\tau \ll 1/\omega_\text{EC}$. In the case, where the coupling to the leads is large, $\gamma_N \gg \omega_\text{CPS}$, and the frequency of elastic cotunneling is small, $\omega_\text{EC} \simeq 0$, we find 
\begin{equation}
g_{l\neq l'}^{(2)}(\tau) \simeq 1-e^{-\gamma_N \tau}\left[\left(1+\frac{\gamma_N \tau}{2}\right)^2-g_x\right].
\end{equation}
Finally, we note that the $g^{(2)}$-correlation functions can be directly related to the noise spectra in Eq.~\eqref{Correlation spectra} as\cite{Emary2012}
\begin{equation}
F_{\ell\ell'} (\omega) =  \delta_{\ell\ell'}+I_N \int_{-\infty}^\infty \text{d}\tau e^{i\omega \tau} \left[ g^{(2)}_{\ell\ell'}(|\tau|)-1\right].
\end{equation}
On the other hand, the charge transport is a non-renewal process, since the system does not return to the same state after each emission event. For this reason, there is no direct connection between the $g^{(2)}$-functions and the distribution of waiting times, and they contain different information about the charge transport statistics.\cite{PhysRevB.99.085418}

\section{Conclusions}\label{Sec:Conclusion}

We have theoretically investigated the noise and the full counting statistics of electrons emitted from a Cooper pair splitter. Working with negatively-biased drain electrodes and a large superconducting gap, the Cooper pair splitter can be described by a Markovian quantum master equation for the dynamics of electrons inside the quantum dots. Using methods from full counting statistics, we have then calculated not only the average current and the shot noise, but also the full frequency-dependent noise spectra, higher-order power-power correlations, as well as the $g^{(2)}$-correlation functions of the output currents. Based on our analytical results for these observables, we have presented a detailed investigation of the fundamental tunneling processes in Cooper pair splitters. Specifically, we have shown how the competing processes of Cooper pair splitting and elastic cotunneling are manifested in the low-frequency fluctuations of the currents and their cross-correlations. If the coupling to the normal-state leads is weak, the two types of processes show up as  dips and peaks in the finite-frequency noise spectrum of the cross-correlations. These results are corroborated by an analysis of the $g^{(2)}$-correlation functions in the time domain. Our work identifies several experimental signatures of the fundamental transport processes in Cooper pair splitters, and we expect that our results may help guide and interpret future experiments on Cooper pair splitting.  

\begin{acknowledgments}
We thank P.~Burset, A.-P.~Jauho, T.~Novotn\'y and M.~Wegewijs for valuable discussions. The Center for Nanostructured Graphene (CNG) is sponsored by the Danish Research Foundation, Project DNRF103. The work was supported by  the  Academy  of  Finland (Projects  No.~308515  and  No.~312299).

\emph{Note added.}--- During the final preparations of our manuscript, we became aware of a related preprint that investigates the current fluctuations in a Cooper pair splitter based on three quantum dots \cite{wrzeniewski2020}.
\end{acknowledgments}

\appendix

\section{Effective Hamiltonian}\label{AppA}

In this appendix, we derive the effective Hamiltonian in Eq.~\eqref{eq:Heff}, assuming a large superconducting gap and strong Coulomb interactions on the quantum dots. In this case, the quantum dots cannot be doubly occupied, and we can discard the double-occupied states in the density matrix and omit the double-occupancy contribution in Eq. \eqref{eq:HQD}.

We start by considering the von Neumann equation for the density matrix of the full system
\begin{equation}\label{eq:vneq}
i\hbar \frac{d}{dt}\hat \rho_H(t)=[\hat H,\hat \rho_H(t)].
\end{equation}
Here $\hat H = \hat H_0 + \hat H_{T_S}$ is the time-independent Hamiltonian, with $\hat H_{T_S}$ the Hamiltonian describing the tunneling between the quantum dots and the superconductor, and $\hat H_0$ is the remaining part of the Hamiltonian. By Laplace-transforming the density matrix as 
\begin{equation}
\hat{\rho}_H(E)=\int_{t_0}^\infty \text{d}t \hat \rho_H(t)e^{i(E+i\eta)(t-t_0)/\hbar},
\end{equation}
we can formally rewrite the von Neumann equation as
\begin{equation}\label{eq:avnlt1}
(E+i\eta)\hat{\rho}_H(E)-i\hbar\hat \rho_H(t_0)=L_0\hat{\rho}_H(E)+L_{T_S}\hat{\rho}_H(E),
\end{equation}
having defined $L_{0/{T_S}}[\ \cdot\ ]=[\hat{H}_{0/{T_S}},\ \cdot\ ]$. We can write the solution as the geometric series
\begin{align}\label{eq:appEHvNGS}
\hat{\rho}_H(E)&=\left(W_0(E)+W_0(E)L_{T_S}W_0(E)\right.\\
&\quad\left.+W_0(E)L_{T_S}W_0(E)L_{T_S}W_0(E)+\cdots\right)\!i\hbar\hat\rho_H(t_0),\nonumber
\end{align}
where $W_0(E)=[E-L_0+i\eta]^{-1}$. The superconductor is in thermal equilibrium, $\hat{\rho}_H(E)\simeq\hat{\rho}_{\tilde{0}}(E)\otimes\hat{\rho}_{SC}^{\text{eq}}$, hence by tracing out the superconductor, we get to second order in $L_{T_S}$,
\begin{equation}
\hat{\rho}_{\tilde{0}}(E)\approx\left(\!W_{\tilde{0}}(E)+W_{\tilde{0}}(E)\text{Tr}_{SC}\!\left[\Sigma_S\hat{\rho}_{SC}^{\text{eq}}\right]\!W_{\tilde{0}}(E)\!\right)\!i\hbar\hat\rho_{\tilde{0}}(t_0),
\end{equation}
where $\hat{H}_{\tilde{0}}=\hat{H}_0-\hat{H}_{SC}$ in $W_{\tilde{0}}$, $\Sigma_S=L_{T_S}W_0(E)L_{T_S}$, and we have used that terms with an odd number of $L_{T_S}$ vanish and that higher-order terms are suppressed in the large gap limit due to $W_0(E)$. Similarly, upon expanding $\hat{\rho}_{\tilde{0}}(E)=(E+i\eta-L_{\tilde{0}}-\overline{\Sigma}_S)^{-1}i\hbar\hat{\rho}_{\tilde{0}}(t_0)$ to first order in $\overline{\Sigma}_S$, we recognize that\cite{Clive:Theory}
$\overline{\Sigma}_S=\text{Tr}_{SC}\left[\Sigma_S\hat{\rho}_{SC}^{\text{eq}}\right]$.

Next, we introduce the Bogoliubov transformation, $\hat{\boldsymbol{\gamma}}_{q}^\dagger=(\hat{\gamma}_{q\uparrow}^\dagger, \hat{\gamma}_{-q\downarrow}^{\phantom\dagger\!})=\hat{\boldsymbol{a}}_{q}^\dagger\boldsymbol{U}_{q}^\dagger$, where $\hat{\boldsymbol{a}}_{q}^\dagger=(\hat{a}_{q\uparrow}^\dagger, \hat{a}_{-q\downarrow}^{\phantom\dagger\!})$ and
\begin{equation}
\boldsymbol{U}_{q}^\dagger = 
\left(
\begin{array}{cc}
u_q^* & v_q \\
-v_q^* & u_q
\end{array}
\right)\!,
\end{equation}
is a unitary matrix with $u_q=(1+\epsilon_q/E_q)^{1/2}/\sqrt{2}$ and $v_q=(1-\epsilon_q/E_q)^{1/2}/\sqrt{2}e^{i\theta_S}$ where $E_q=\sqrt{\epsilon_q^2+|\Delta|^2}$ and $\theta_S$ is the phase of the superconductor. With this transformation, we get $\hat{H}_{SC}=\sum_{q\sigma}E_q\hat{\gamma}_{q\sigma}^\dagger\hat{\gamma}_{q\sigma}^{\phantom\dagger\!}$ (plus a constant, which does not contribute to the von Neumann equation), and the tunneling Hamiltonian \eqref{eq:HTS} becomes
\begin{equation}
\hat{H}_{T_S}=\sum_{\xi=\pm,\ell q\sigma}\xi t_{S\ell q}^{\xi}\left(u_q^{\xi}\hat{\gamma}_{q\sigma}^{\xi}+\sigma v_q^{(-\xi)}\hat{\gamma}_{-q-\sigma}^{(-\xi)}\right)\hat{d}_{\ell\sigma}^{\xi},
\end{equation}
where we have defined $t_{S\ell q}^{(+)-}=t_{S\ell q}^{(\ )*}$, $u_{q}^{(+)-}=u_{q}^{(\ )*}$, $v_{q}^{(+)-}=v_{q}^{(\ )*}$, $\hat{\gamma}_{q\sigma}^{+(-)}=\hat{\gamma}_{q\sigma}^{\dagger(\ )}$, $\hat{d}_{\ell\sigma}^{(+)-}=\hat{d}_{\ell\sigma}^{(\ )\dagger}$. We can furthermore write $L_{T_S}$ in the compact form\cite{Clive:Theory}
\begin{equation}
L_{T_S}=\!\!\sum_{\xi,\theta=\pm,\ell q\sigma}\!\!\xi t_{S\ell q}^{\xi}\left(u_q^\xi\Gamma_{q\sigma}^{\xi\theta}+\sigma v_{q}^{(-\xi)}\Gamma_{-q-\sigma}^{(-\xi)\theta}\right)D_{\ell \sigma}^{\xi\theta},
\end{equation}
where $\theta=\pm$ determines if the operator acts to the left ($+$) or right ($-$), for instance
\begin{equation}
\Gamma_{q \sigma}^{\xi+}\hat{O}=\hat{\gamma}_{q \sigma}^{\xi}\hat{O}, \quad \Gamma_{q \sigma}^{\xi-}\hat{O}=\hat{O}\hat{\gamma}_{q\sigma}^{\xi},
\end{equation} 
and
\begin{equation}
D_{\ell \sigma}^{\xi+}\hat{O}=\hat{d}_{\ell \sigma}^{\xi}\hat{O}, \quad D_{\ell \sigma}^{\xi-}\hat{O}=\hat{O}\hat{d}_{\ell\sigma}^{\xi},
\end{equation}
where $\hat{O}$ is an operator. With these transformations, we readily obtain
\begin{widetext}
\begin{equation}
\Sigma_S = 
\!\!\sum_{\xi\theta\ell q\sigma\phantom'\!}\sum_{\xi'\!\theta'\!\ell'\! q'\!\sigma'\!}\!\!\xi\xi' t_{S\ell q}^{\xi}t_{S\ell' q'}^{\xi'}D_{\ell'\sigma'}^{\xi'\theta'}\left(u_{q'}^{\xi'}\Gamma_{q'\sigma'}^{\xi'\theta'}+\sigma' v_{q'}^{(-\xi')}\Gamma_{-q'-\sigma'}^{(-\xi')\theta'}\right)
W_{0}(E)
D_{\ell \sigma}^{\xi\theta}\left(u_q^\xi\Gamma_{q\sigma}^{\xi\theta}+\sigma v_{q}^{(-\xi)}\Gamma_{-q-\sigma}^{(-\xi)\theta}\right)\!,
\end{equation}
where we have used the commutation relation $\Gamma^{\chi\theta}D^{\chi'\theta'}=-\theta\theta'D^{\chi'\theta'}\Gamma^{\chi\theta}$ (suppressing the subscripts). Having expressed the tunneling Hamiltonian in terms of the Bogoliubov transformation that diagonalizes the superconducting Hamiltonian, we have
\begin{equation}
\Gamma_{q\sigma}^{\xi\theta}L_0=(L_0-\xi E_q)\Gamma_{q\sigma}^{\xi\theta},
\end{equation}
and thus
\begin{equation}
\begin{split}
\Sigma_S &= 
-\!\!\sum\!\xi\xi'\theta\theta' t_{S\ell q}^{\xi}t_{S\ell' q'}^{\xi'}\!D_{\ell'\sigma'}^{\xi'\theta'}\!\left(\!W_{0}(E+\xi'\!E_{q'}\!)D_{\ell \sigma}^{\xi\theta}u_{q'}^{\xi'}\Gamma_{q'\sigma'}^{\xi'\theta'}+W_{0}(E-\xi'\!E_{q'}\!)D_{\ell \sigma}^{\xi\theta}\sigma'\! v_{q'}^{(-\xi')}\Gamma_{-q'-\sigma'}^{(-\xi')\theta'}\!\right)\\
&\quad\times
\left(\!u_q^\xi\Gamma_{q\sigma}^{\xi\theta}+\sigma v_{q}^{(-\xi)}\Gamma_{-q-\sigma}^{(-\xi)\theta}\right)\!,
\end{split}
\end{equation}
where we have left the summation indices implicit. Upon tracing out the superconductor, we find
\begin{equation}\label{eq:AppAS3}
\overline{\Sigma}_S=\!\!\sum_{\xi\theta\theta'\ell\ell'\sigma}\!\!\!\!\theta\theta'\!\left(D_{\ell'\sigma}^{(-\xi)\theta'}I^{(1)}_{\xi\theta\ell\ell'}+\sigma D_{\ell'-\sigma}^{\xi\theta'}I^{(2)}_{\xi\theta\ell\ell'}\right)\!D_{\ell\sigma}^{\xi\theta},
\end{equation}
where we have defined
\begin{equation}
I^{(1)}_{\xi\theta\ell\ell'}=\sum_{q}t_{S\ell q}^{\xi}t_{S\ell'q}^{(-\xi)}\left(|u_q|^2f^{(-\xi\theta)}(E_q)W_{\tilde{0}}(E-\xi E_q)+|v_q|^2f^{(\xi\theta)}(E_q)W_{\tilde{0}}(E+\xi E_q)\right),\label{eq:AppAI1}
\end{equation}
and
\begin{equation}
I^{(2)}_{\xi\theta\ell\ell'}=\sum_{q}t_{S\ell q}^{\xi}t_{S\ell'-q}^{\xi}u_q^{\xi}v_q^{(-\xi)}\left(f^{(-\xi\theta)}(E_q)W_{\tilde{0}}(E-\xi E_q)-f^{(\xi\theta)}(E_q)W_{\tilde{0}}(E+\xi E_q)\right),\label{eq:AppAI2}
\end{equation}
and we have used that $\text{Tr}_{SC}\left[\Gamma_{q'\sigma'}^{\xi'\theta'}\Gamma_{q\phantom'\!\sigma\phantom'\!}^{\xi\phantom'\!\theta\phantom'\!}\hat{\rho}_{SC}^{\text{eq}}\right]=\delta_{qq'}\delta_{\sigma\sigma'}\delta_{\xi,-\xi'}f^{(-\xi\theta)}(E_q)$, $f^+=f$ and $f^-=1-f$ with $f$ being the Fermi--Dirac distribution, and $\epsilon_{q}=\epsilon_{-q}$. In the limit of large superconducting gap at long times, $W_{\tilde{0}}(E\pm\xi E_q)$ is dominated by the constant factor $\pm\xi E_q^{-1}$, whereby
\begin{equation}
I^{(1)}_{\xi\theta\ell\ell'}\simeq-\sum_{q}t_{S\ell q}^{\xi}t_{S\ell' q}^{(-\xi)}\xi E_{q}^{-1}\!\left(|u_q|^2f^{(-\xi\theta)}(E_q)-|v_q|^2f^{(\xi\theta)}(E_q)\right),
\end{equation}
and
\begin{equation}
I^{(2)}_{\xi\theta\ell\ell'}\simeq I^{(2)}_{\xi\ell\ell'}=-\sum_{q}t_{S\ell q}^{\xi}t_{S\ell'-q}^{\xi}u_q^\xi v_q^{(-\xi)}\xi E_q^{-1}.
\end{equation}
\end{widetext}
Using that $I^{(1)}_{(-\xi)\theta\ell'\ell}=-I^{(1)}_{\xi(-\theta)\ell\ell'}$ and $I^{(2)}_{\xi\ell'\ell}=I^{(2)}_{\xi\ell\ell'}$ we find upon performing the sum over $\theta$ and $\theta'$ in Eq.~\eqref{eq:AppAS3} $\overline{\Sigma}_S[\ \cdot\ ]=[\hat{H}_{\Sigma_S},\ \cdot\ ]$, where
\begin{equation}
\hat{H}_{\Sigma_S}=\sum_{\xi\ell\ell'\sigma}\left(I^{(1)}_{\xi+\ell\ell'}\hat{d}^{(-\xi)}_{\ell'\sigma}\hat{d}^{\xi}_{\ell\sigma}+\sigma I^{(2)}_{\xi\ell\ell'}\hat{d}^{\xi}_{\ell'-\sigma}\hat{d}^{\xi}_{\ell'\sigma}\right).
\end{equation}

Carrying out the remaining sums, one obtains the terms in Eq.~\eqref{eq:Heff}, where we have defined the amplitudes $\hbar\gamma_\text{CPS}=-\sqrt{2}(I^{(2)}_{-LR}+I^{(2)}_{-RL})$ and $\hbar\gamma_\text{EC}=I^{(1)}_{-+LR}-I^{(1)}_{++RL}$, corresponding to Cooper pair splitting and elastic cotunneling, respectively, absorbed the constant self-energy into a redefinition of the quantum dot levels, and omitted the term corresponding to a Cooper pair occupying a single dot, which is prevented in the large-$U$ limit. The momentum integrals from $I^{(1)}_{\xi\theta\ell\ell'}$ and $I^{(2)}_{\xi\ell\ell'}$ are calculated explicitly in Ref.~\onlinecite{Sauret:Quantum} assuming point-like contacts between each dot and the superconductor (with zero temperature), separated by the distance $\delta r$.

\section{Quantum master equation}\label{AppDiss}

In this appendix, we derive the quantum master equation~\eqref{eq:vonNeumannEq} with the dissipator given by Eq.~\eqref{eq:Dissipator}. A microscopic approach for quantum transport in normal-state structures in the high-bias limit has been devised by Gurvitz and Prager \cite{Gurvitz_1996,Gurvitz:Rate,Gurvitz:Wave} and later on extended to Cooper pair splitters in Ref.~\onlinecite{Sauret:Quantum}. The method uses an occupation-number representation of the many-body wave function, which is time-evolved under the Schr\"odinger equation. As an alternative and potentially more compact approach, we here derive the quantum master equation starting from the von Neumann equation for $\hat{\rho}_{\tilde{0}}$ (cf. App.~\ref{AppA}).
\iffalse
\begin{equation}\label{eq:vneq}
i\hbar \frac{d}{dt}\hat \rho(t)=[\hat H,\hat \rho(t)].
\end{equation}
Here $\hat H = \hat H_0 + \hat H_{T_N}$ is the time-independent Hamiltonian, with $\hat H_0$ the Hamiltonian of the uncoupled system and $\hat H_{T_N}$ the Hamiltonian describing the tunneling between the dots and the normal-state electrodes. By Laplace-transforming the density matrix as 
\begin{equation}
\hat{\rho}(E)=\int_{t_0}^\infty \text{d}t \hat \rho(t)e^{i(E+i\eta)(t-t_0)},
\end{equation}
we can formally rewrite the von Neumann equation as
\begin{equation}\label{eq:vnlt1}
(E+i\eta)\hat{\rho}(E)-i\hat \rho(t_0)=L_0\hat{\rho}(E)+L_{T_N}\hat{\rho}(E),
\end{equation}
having defined $L_{0/{T_N}}[\ \cdot\ ]=\frac{1}{\hbar}[H_{0/{T_N}},\ \cdot\ ]$.

We can now write the solution as the geometric series
\begin{equation}
\hat{\rho}(E)=\left(W_0(E)+W_0(E)L_{T_N}W_0(E)+\cdots\right)i\hat\rho(t_0),
\end{equation}
where $W_0(E)=[E-L_0+i\eta]^{-1}$ is the free propagator.
\fi
The geometric series form of the von Neumann equation as in Eq.~\eqref{eq:appEHvNGS} can also be obtained by iterating as
\begin{align}\label{eq:vnLTIt}
\hat{\rho}_{\tilde{0}}(E)&=W_{\bar{0}}(E)\left[L_{T_N}\hat{\rho}_{\tilde{0}}(E)+i\hbar\hat{\rho}_{\tilde{0}}(t_0)\right]\\
&=W_{\bar{0}}(E)\left[L_{T_N}W_{\bar{0}}(E)\left(L_{T_N}\hat{\rho}_{\tilde{0}}(E)+i\hbar\rho_{\tilde{0}}(t_0)\right)\right.\nonumber\\
&\quad\left.+i\hbar\hat\rho_{\tilde{0}}(t_0)\right]\nonumber\\
&=\cdots\nonumber,
\end{align}
where $\hat{H}_{\bar{0}}=\hat{H}_{S}+\hat{H}_{N}$ in $W_{\bar{0}}$. We now inspect the operator 
\begin{equation}
\Sigma_N\equiv L_{T_N}W_{\bar{0}}(E)L_{T_N},
\end{equation} 
which appears after the first iteration. To this end, we express the tunneling Hamiltonian in the compact form
\begin{equation}
\hat H_{T_N}=\sum_{\xi=\pm,\ell k\sigma}\xi t_{\ell k}^{\xi}\hat{c}_{\ell k\sigma}^{\xi} \hat{d}_{\ell\sigma}^{\xi},
\end{equation}
where we have defined $\hat{c}_{\ell k\sigma}^{+(-)}=\hat{c}_{\ell k \sigma}^{\dagger(\ )}$, $\hat{d}_{\ell\sigma}^{+(-)}=\hat{d}_{\ell\sigma}^{(\ )\dagger}$, and $t_{\ell k}^+=t_{\ell k}$, $t_{\ell k}^-=t_{\ell k}^*$. We also have
\begin{equation}
L_{T_N}=\sum_{\xi,\theta=\pm,\ell k\sigma}\xi t_{\ell k}^{\xi}C_{\ell k \sigma}^{\xi\theta}D_{\ell \sigma}^{\xi\theta},
\end{equation}
in terms of superoperators as in App.~\ref{AppA}.
With these definitions, we readily obtain 
\begin{equation}
\begin{split}
\Sigma_N&=\!\sum_{\xi\theta\ell k\sigma\phantom'}\!\sum_{\xi'\theta'\ell' k'\sigma'}\!\!\!\xi\xi't_{\ell k}^{\xi}t_{\ell'k'}^{\xi'}D_{\ell'\sigma'}^{\xi'\theta'}C_{\ell'k'\sigma'}^{\xi'\theta'}W_{\bar{0}}(E)D_{\ell\sigma}^{\xi\theta}C_{\ell k\sigma}^{\xi\theta},
\end{split}
\end{equation}
having used the commutation relation  $C^{\xi\theta}D^{\xi'\theta'}=-\theta\theta'D^{\xi'\theta'}C^{\xi\theta}$, omitting the subscripts here. 

The electrons in the normal-state reservoirs are non-interacting, such that
\begin{equation}
C_{\ell k\sigma}^{\xi\theta}L_{\bar{0}}=(L_{\bar{0}}-\xi\epsilon_{\ell k})C_{\ell k\sigma}^{\xi\theta},
\end{equation} 
and thus
\begin{align}
\Sigma_N&=-\sum t_{\ell k}^{\xi}t_{\ell'k'}^{\xi'}D_{\ell'\sigma'}^{\xi'\theta'}W_{\bar{0}}(E+\xi'\epsilon_{\ell'k'})D_{\ell\sigma}^{\xi\theta}C_{\ell'k'\sigma'}^{\xi'\theta'}C_{\ell k\sigma}^{\xi\theta}\nonumber\\
&\quad\times \xi\xi'\theta\theta',
\end{align}
where we have left out the summation indices in the sum. The environment is not affected by the subsystem of interest, $\hat{\rho}_{\tilde{0}}(E)\simeq\hat{\rho}(E)\otimes\hat{\rho}_N^{\text{eq}}$, allowing us to trace out the environmental degrees of freedom as
\begin{align}
\hat{\overline{\Sigma}}_\rho&=\!\!\!\!\sum_{\xi\theta\theta'\ell k\sigma}\!\!\!\!\theta\theta't_{\ell k}^{\xi}t_{\ell k}^{-\xi}D_{\ell\sigma}^{(-\xi)\theta'}W_S(E-\xi\epsilon_{\ell k})D_{\ell\sigma}^{\xi\theta}\hat{\rho}(E)\nonumber\\
&\quad\times f_{\ell}^{(-\xi\theta)}(\epsilon_{\ell k}),
\end{align}
where $\hat{H}_S$ in $W_S$ is given in Eq.~\eqref{eq:Heff}, we have defined $\hat{\overline{\Sigma}}_\rho=\text{Tr}_{N}\left\{\Sigma_N\hat{\rho}_{\tilde{0}}(E)\right\}$ and used that $\text{Tr}_N\{C_{\ell'k'\sigma'}^{\xi'\theta'}C_{\ell k \sigma}^{\xi\theta}\hat{\rho}_N^{\text{eq}}\}=\delta_{\ell\ell'}\delta_{kk'}\delta_{\sigma\sigma'}\delta_{\xi,-\xi'}f_{\ell}^{(-\xi\theta)}(\epsilon_{\ell k})$, where $f^+_{\ell}=f_{\ell}$ is the Fermi--Dirac distribution, and $f^-_{\ell}=1-f_{\ell}$. Formally, inserting completeness relations in terms of the eigenstates of $\hat{H}_S=\sum_{a}\epsilon_a\ket{a}\bra{a}$, we find
\begin{equation}
\label{eq:completeness}
\hat{\overline{\Sigma}}_\rho=\!\!\!\!\sum_{\xi\theta\theta'\ell \sigma}\!\sum_{aa'}\theta\theta'D_{\ell\sigma}^{(-\xi)\theta'}\ket{a}\bra{a}(D_{\ell\sigma}^{\xi\theta}\hat{\rho}(E))\ket{a'}\bra{a'}I_{\delta\delta'\xi\theta aa'},
\end{equation}
where the integral
\begin{equation}
\label{eq:integral}
I_{\ell\xi\theta aa'}=\int\!d\epsilon\frac{\nu_{\ell}(\epsilon)|t_{\ell}(\epsilon)|^2f_{\ell}^{(-\xi\theta)}(\epsilon)}{E-\xi\epsilon+i\eta-(\epsilon_a-\epsilon_{a'})},
\end{equation}
contains the density of states, $\nu_{\ell}$, of lead $\ell$. 

We now assume that large negative voltages are applied to the normal-state electrodes, so that they are completely empty, $f_{\ell}^{(-\xi\theta)}(\epsilon)=\delta_{-\xi\theta,-}$. In addition, we assume that the lead coupling, 
\begin{equation}
\gamma_{\ell}\equiv \frac{2\pi}{\hbar}\nu_{\ell}|t_{\ell}|^2,
\end{equation} 
is constant for the relevant energies. Hence,
\begin{equation}
I_{\ell\xi\theta aa'}=-i\hbar\gamma_{\ell}\delta_{-\xi\theta,-}/2\equiv -i\hbar I_{\ell\xi\theta}/2.
\end{equation} 
Since $I_{\ell\xi\theta}$ does not depend on $a$ and $a'$ we can remove the completeness relations from Eq.~\eqref{eq:completeness} and write
\begin{equation}
\hat{\overline{\Sigma}}_\rho=-\frac{i\hbar}{2}\sum_{\xi\theta\theta'\ell\sigma}\!\!\theta\theta'D_{\ell\sigma}^{(-\xi)\theta'}D_{\ell\sigma}^{\xi\theta}I_{\ell\xi\theta}\hat{\rho}(E).
\end{equation}

Considering next the following iterations in Eq.~\eqref{eq:vnLTIt}, we see that terms with an odd number of $L_{T_N}$ vanish, once we trace out the environment. On the other hand, for terms with an even number of $L_{T_N}$, we see that as we commute all the $C$'s to the right, the leftmost $C$ will give rise to the substitution, $E\to E+\xi\epsilon$, in all the $W_{\bar{0}}$'s. Hence, the approximations used above lead to integrals over $\epsilon$, as in Eq.~(\ref{eq:integral}), involving products of simple fractions with poles in the same complex half-plane. For this reason, these integrals vanish \cite{Gurvitz_1996}. As a result, the iteration loop terminates, and upon tracing out the environment, we can write Eq.~\eqref{eq:vnLTIt} as
\begin{equation}
(E+i\eta-L_S)\hat{\rho}(E)=i\hbar\mathcal{D}\hat{\rho}(E)+i\hbar\hat \rho(t_0),
\end{equation}
where we have defined the superoperator
\begin{equation}
\mathcal{D}=-\frac{1}{2}\!\!\sum_{\xi\theta\theta'\ell\sigma}\theta\theta'D_{\ell\sigma}^{(-\xi)\theta'}D_{\ell\sigma}^{\xi\theta}I_{\ell\xi\theta}.
\end{equation}
Finally, by transforming this expression back to the time domain, we arrive at Eqs.~\eqref{eq:vonNeumannEq} and \eqref{eq:Dissipator} by letting the dissipator $\mathcal D$ act on the reduced density matrix.

\begin{widetext}
\section{Matrix representation}\label{AppC}
To carry out our calculations, we need a matrix representation of the Liouvillian. In the basis 
\begin{equation}
\{\rho_{(0)(0)},\rho_{(L\uparrow)(L\uparrow)},\rho_{(L\downarrow)(L\downarrow)},\rho_{(R\uparrow)(R\uparrow)},
\rho_{(R\downarrow)(R\downarrow)},\rho_{(S)(S)}, \rho_{(0)(S)},\rho_{(S)(0)},\rho_{(L\uparrow)(R\uparrow)}, \rho_{(R\uparrow)(L\uparrow)},\rho_{(L\downarrow)(R\downarrow)}, \rho_{(R\downarrow)(L\downarrow)}\},
\end{equation}
where $\rho_{\psi\psi'}=\bra{\psi}\hat \rho\ket{\psi'}$ and $\ket{\psi},\ket{\psi'}\in\{\ket{0},\ket{\ell\sigma}=\hat{d}_{\ell\sigma}^\dagger\ket{0},\ket{S}=\frac{1}{\sqrt{2}}\!\left(\!\hat{d}_{L\downarrow}^\dagger \hat{d}_{R\uparrow}^\dagger-\hat{d}_{L\uparrow}^\dagger \hat{d}_{R\downarrow}^\dagger\!\right)\!\ket{0}\}$, the Liouvillian reads
\begin{equation}\label{eq:kernelmatrix1}
\mathcal{L}\!=\!\!
\left( 
\scalemath{0.85}{
\begin{array}{cccccccccccc}
0 & \gamma_Le^{i\chi_{L}} & \gamma_Le^{i\chi_{L}} & \gamma_Re^{i\chi_{R}} & \gamma_Re^{i\chi_{R}} & 0 & -i\gamma_\text{CPS} & i\gamma_\text{CPS} & 0 & 0 & 0 & 0 \\
0 & -\gamma_L & 0 & 0 & 0 & \frac{\gamma_R}{2}e^{i\chi_{R}} & 0 & 0 &  -i\gamma_\text{EC} &  i\gamma_\text{EC} & 0 & 0 \\
0 & 0 & -\gamma_L & 0 & 0 & \frac{\gamma_R}{2}e^{i\chi_{R}} & 0 & 0 & 0 & 0 & -i\gamma_\text{EC} & i\gamma_\text{EC} \\
0 & 0 & 0 & -\gamma_R & 0 & \frac{\gamma_L}{2}e^{i\chi_{L}} & 0 & 0 & i\gamma_\text{EC} & -i\gamma_\text{EC} & 0 & 0 \\
0 & 0 & 0 & 0 & -\gamma_R & \frac{\gamma_L}{2}e^{i\chi_{L}} & 0 & 0 & 0 & 0 & i\gamma_\text{EC} & -i\gamma_\text{EC} \\
0 & 0 & 0 & 0 & 0 & -(\gamma_L\!+\!\gamma_R) & i\gamma_\text{CPS} & -i\gamma_\text{CPS} & 0 & 0 & 0 & 0 \\
-i\gamma_\text{CPS} & 0 & 0 & 0 & 0 & i\gamma_\text{CPS} & i\varepsilon\!-\!\frac{\gamma_L\!+\!\gamma_R}{2} & 0 & 0 & 0 & 0 & 0 \\
i\gamma_\text{CPS} & 0 & 0 & 0 & 0 & -i\gamma_\text{CPS} & 0 & -i\varepsilon\!-\!\frac{\gamma_L\!+\!\gamma_R}{2} & 0 & 0 & 0 & 0 \\
0 & -i\gamma_\text{EC} & 0 & i\gamma_\text{EC} & 0 & 0 & 0 & 0 & -i\delta\!-\!\frac{\gamma_L\!+\!\gamma_R}{2} & 0 & 0 & 0 \\
0 & i\gamma_\text{EC} & 0 & -i\gamma_\text{EC} & 0 & 0 & 0 & 0 & 0 & i\delta\!-\!\frac{\gamma_L\!+\!\gamma_R}{2} & 0 & 0\\
0 & 0 & -i\gamma_\text{EC} & 0 & i\gamma_\text{EC} & 0 & 0 & 0 & 0 & 0 & -i\delta\!-\!\frac{\gamma_L\!+\!\gamma_R}{2} & 0\\
0 & 0 & i\gamma_\text{EC} & 0 & -i\gamma_\text{EC} & 0 & 0 & 0 & 0 & 0 & 0 & i\delta\!-\!\frac{\gamma_L\!+\!\gamma_R}{2}\\
\end{array}
}
\right)\!\!,
\end{equation}
where we have introduced 
the counting fields $\chi_L$ and $\chi_R$ that couple to transitions into the left and right leads.
\end{widetext}

\section{Power-power correlations}\label{AppD}
To evaluate the cumulants of the full counting statistics, we need to find the derivatives of the eigenvalue $\lambda_0(\boldsymbol{\chi})$ of $\mathcal{L}(\boldsymbol{\chi})$  with the largest real part. For $\boldsymbol{\chi}=\boldsymbol{0}$, this is the zero-eigenvalue, $\lambda_0(\boldsymbol{0})=0$, corresponding to the stationary state $\hat\rho_S$, defined as the normalized solution to $\mathcal{L}\hat\rho_S=0$, which constitutes our unperturbed problem.

We now follow Refs.~\onlinecite{Flindt2005,Flindt2008,Flindt2010} and calculate $\lambda_0(\boldsymbol{\chi})$ perturbatively in the counting fields, $\boldsymbol{\chi}=(\chi_{L},\chi_{R})$.  Our starting point is the perturbed eigenvalue problem
\begin{equation}
\mathcal{L}(\boldsymbol{\chi})\hat\rho_S(\boldsymbol{\chi}) =  [\mathcal{L}+\mathcal L'(\boldsymbol{\chi})]\hat\rho_S(\boldsymbol{\chi})=\lambda_0(\boldsymbol{\chi})\hat\rho_S(\boldsymbol{\chi}),
\end{equation}
where $\mathcal L'(\boldsymbol{\chi})$ is the perturbation due to the counting fields. Following the steps of Refs.~\onlinecite{Flindt2005,Flindt2008,Flindt2010}, we find
\begin{equation}
\lambda_0(\boldsymbol{\chi}) = \mathrm{Tr} \left\{\mathcal L'(\boldsymbol{\chi})\hat\rho_S(\boldsymbol{\chi})\right\},
\label{lambda eq}
\end{equation}
and
\begin{equation}
\hat\rho_S(\boldsymbol{\chi}) = \hat\rho_S+\mathcal{R}[\lambda_0(\boldsymbol{\chi})-\mathcal L'(\boldsymbol{\chi})]\hat\rho_S(\boldsymbol{\chi}),
\label{0 exp eq}
\end{equation}
where $\mathcal{R} = \mathcal{R}(0)$ is the pseudo-inverse in Eq.~(\ref{eq_pseudo}) evaluated at $\omega=0$. Next, we expand all quantities as
\begin{equation}
\begin{split}
\lambda_0(\boldsymbol{\chi}) &= \sum_{n,m=0}^\infty\frac{(i\chi_L)^n}{n!}\frac{(i\chi_R)^m}{m!}\langle\!\langle I_L^nI_R^m\rangle\!\rangle,\\
\hat\rho_S(\boldsymbol{\chi}) &= \sum_{n,m=0}^\infty \frac{(i\chi_L)^n}{n!}\frac{(i\chi_R)^m}{m!}\hat\rho_S^{(n,m)},\\
\mathcal{L}'(\boldsymbol{\chi}) &= \sum_{n,m=0}^\infty \frac{(i\chi_L)^n}{n!}\frac{(i\chi_R)^m}{m!} \mathcal{L}^{(n,m)},
\end{split}
\end{equation}
recalling that the cumulants of the currents are given by the derivatives of the largest eigenvalue.  We also note that $\mathcal{L}^{(0,0)} = \mathcal L'(\boldsymbol{0})= 0$ by definition.

Inserting these expansions into Eqs.~\eqref{lambda eq} and \eqref{0 exp eq} and collecting terms to same order in the counting fields, we obtain the recursive formulas~\cite{Flindt2008,Flindt2010}
\begin{equation}
\langle\!\langle I_L^nI_R^m\rangle\!\rangle = \sum_{i,j=0}^{n,m} \begin{pmatrix}
n \\
i
\end{pmatrix}\!\!
\begin{pmatrix}
m \\
j
\end{pmatrix}
\mathrm{Tr} \left\{\mathcal{L}^{(i,j)}\hat\rho_S^{(n-i,m-j)}\right\},
\label{c formula}
\end{equation}
and
\begin{equation}
\hat\rho_S^{(n,m)} = \mathcal{R}\sum_{i,j=0}^{n,m} \begin{pmatrix}
n \\
i
\end{pmatrix}\!\!
\begin{pmatrix}
m \\
j
\end{pmatrix}
\left[\langle\!\langle I_L^iI_R^j\rangle\!\rangle-\mathcal{L}^{(i,j)}\right]\hat\rho_S^{(n-i,m-j)}.
\label{0 formula}
\end{equation}
From these expressions, we can in principle calculate any cumulant of the currents. For the Cooper pair splitter, the calculations are simplified by the fact that $\mathcal{L}^{(n,m)} = 0$, if both $n>0$ and $m>0$. As an illustration of the recursive scheme, we find for some of the corrections to the eigenstate the following expressions
\begin{equation}
\begin{split}
\hat\rho_S^{(0,1)} =& -\mathcal{R} \mathcal{J}_R\hat\rho_S,\\
\hat\rho_S^{(0,2)}= &-\mathcal{R}\left[2\mathcal I_R\mathcal{R} +1\right]\mathcal J_R\hat\rho_S,\\
\hat\rho_S^{(1,1)} =& -\mathcal{R} \left[ \mathcal I_R\mathcal{R}\mathcal J_L +\mathcal I_L\mathcal{R} \mathcal J_R\right] \hat\rho_S,\\
\hat\rho_S^{(1,2)} =&-\mathcal{R} \big[  2\mathcal I_R\mathcal{R} \left[  \mathcal I_R\mathcal{R} \mathcal J_L + \mathcal I_L\mathcal{R} \mathcal J_R \right] + \mathcal S_{R}\mathcal{R} \mathcal J_L\\
&+\mathcal I_L \mathcal{R}\left[2 \mathcal I_R\mathcal{R}+1\right] \mathcal J_R+ 2\langle\!\langle I_LI_R\rangle\!\rangle\mathcal{R} \mathcal J_R  \big]\hat\rho_S,
\end{split}
\end{equation}
where $\mathcal{I}_{\ell} = \langle I_{\ell}\rangle-\mathcal J_{\ell}$ and $\mathcal S_{\ell} = \langle \! \langle  I^2_{\ell}\rangle\!\rangle-\mathcal{J}_\ell$ is given in terms of the noise, $\langle \! \langle  I^2_{\ell}\rangle\!\rangle=
\text{Tr}\big[\mathcal{J}_\ell\hat\rho_S\big]-2\text{Tr}\big[  \mathcal{J}_\ell\mathcal{R}\mathcal{J}_{\ell}\hat\rho_S\big]=S_{\ell\ell}(0)$.
Based on these expressions and Eq.~\eqref{c formula}, we obtain Eqs.~(\ref{Power-power auto}) and~(\ref{Power-power correlation}) for the power-power correlations.

%\bibliography{journalabbreviations,references}

% For submission
%\bibliographystyle{apsrev}
%\bibliography{paper.bbl}

%

\end{document}